\documentclass[aps,pra,reprint,groupedaddress]{revtex4-1}

\usepackage{amssymb}

\usepackage{lipsum}
\usepackage[figuresright]{rotating}
\usepackage{graphicx}
\usepackage{textcomp}

\usepackage{titlesec}

\begin{document}





\title{Atomic fountains and optical clocks at SYRTE: status and perspectives}

\author{M.~Abgrall}
\affiliation{LNE-SYRTE, Observatoire de Paris, PSL Research University, CNRS, Sorbonne Universit\'{e}s, UPMC University Paris 06, 61, avenue de l$\,'\!$Observatoire, 75014 Paris, France}
\author{B.~Chupin}
\affiliation{LNE-SYRTE, Observatoire de Paris, PSL Research University, CNRS, Sorbonne Universit\'{e}s, UPMC University Paris 06, 61, avenue de l$\,'\!$Observatoire, 75014 Paris, France}\author{L.~De~Sarlo}
\affiliation{LNE-SYRTE, Observatoire de Paris, PSL Research University, CNRS, Sorbonne Universit\'{e}s, UPMC University Paris 06, 61, avenue de l$\,'\!$Observatoire, 75014 Paris, France}
\author{J.~Gu\'ena}
\affiliation{LNE-SYRTE, Observatoire de Paris, PSL Research University, CNRS, Sorbonne Universit\'{e}s, UPMC University Paris 06, 61, avenue de l$\,'\!$Observatoire, 75014 Paris, France}
\author{Ph.~Laurent}
\affiliation{LNE-SYRTE, Observatoire de Paris, PSL Research University, CNRS, Sorbonne Universit\'{e}s, UPMC University Paris 06, 61, avenue de l$\,'\!$Observatoire, 75014 Paris, France}
\author{Y.~Le~Coq}
\affiliation{LNE-SYRTE, Observatoire de Paris, PSL Research University, CNRS, Sorbonne Universit\'{e}s, UPMC University Paris 06, 61, avenue de l$\,'\!$Observatoire, 75014 Paris, France}
\author{R.~Le~Targat}
\affiliation{LNE-SYRTE, Observatoire de Paris, PSL Research University, CNRS, Sorbonne Universit\'{e}s, UPMC University Paris 06, 61, avenue de l$\,'\!$Observatoire, 75014 Paris, France}
\author{J.~Lodewyck}
\affiliation{LNE-SYRTE, Observatoire de Paris, PSL Research University, CNRS, Sorbonne Universit\'{e}s, UPMC University Paris 06, 61, avenue de l$\,'\!$Observatoire, 75014 Paris, France}
\author{M.~Lours}
\affiliation{LNE-SYRTE, Observatoire de Paris, PSL Research University, CNRS, Sorbonne Universit\'{e}s, UPMC University Paris 06, 61, avenue de l$\,'\!$Observatoire, 75014 Paris, France}
\author{P.~Rosenbusch}
\affiliation{LNE-SYRTE, Observatoire de Paris, PSL Research University, CNRS, Sorbonne Universit\'{e}s, UPMC University Paris 06, 61, avenue de l$\,'\!$Observatoire, 75014 Paris, France}
\author{D.~Rovera}
\affiliation{LNE-SYRTE, Observatoire de Paris, PSL Research University, CNRS, Sorbonne Universit\'{e}s, UPMC University Paris 06, 61, avenue de l$\,'\!$Observatoire, 75014 Paris, France}
\author{S.~Bize}
\email{sebastien.bize@obspm.fr}
\affiliation{LNE-SYRTE, Observatoire de Paris, PSL Research University, CNRS, Sorbonne Universit\'{e}s, UPMC University Paris 06, 61, avenue de l$\,'\!$Observatoire, 75014 Paris, France}


\begin{abstract}

In this article, we report on the work done with the LNE-SYRTE atomic clock ensemble during the last 10 years. We cover progress made in atomic fountains and in their application to timekeeping. We also cover the development of optical lattice clocks based on strontium and on mercury. We report on tests of fundamental physical laws made with these highly accurate atomic clocks. We also report on work relevant to a future possible redefinition of the SI second.

\end{abstract}

\maketitle

%
%
%
%



\section*{Introduction}

Research on highly accurate atomic frequency standards and their applications is making fast and steady progress. The quest for ever increased accuracy is advancing hand in hand with advances in quantum physics, with better understanding and manipulation of atomic systems, with exploration of fundamental laws of nature and with the development of important services and infrastructures for science and society. The quest for increased accuracy is also a powerful incentive to innovation in such areas as lasers, laser stabilization, low noise electronics, stable oscillators, low noise detection of optical signals, fiber devices and cold-atom based instrumentation for ground or space applications. Finally, in addition to enhancing existing applications, improved accuracy leads to new applications.

This article focuses on key achievements and trends of the last 10 years. Over this period of time, the first generation of laser-cooled standards, using the atomic fountain geometry, reached maturity and had large impact on international timekeeping. The typical accuracy of these frequency standards,  2 parts in $10^{16}$, is now permanently accessible both locally and globally. At the same time, a new generation of optical clocks showed tremendous and steady improvement, gaining more than two orders of magnitude in a decade. To date, an accuracy of $6.4\times 10^{-18}$ \cite{bloom2014optical} was reported. Similarly, major improvements occurred in many other aspects of optical frequency metrology, notably in optical frequency combs and optical fiber links.

In this article, we will report on developments of the LNE-SYRTE atomic clock ensemble since our 2004 report in the Special Issue of the Comptes Rendus de l'Acad\'emie des sciences on Fundamental Metrology \cite{Bize2004}. This work exemplifies many of the above mentioned features of research on highly accurate atomic frequency standards. We will focus on frequency standards and their impact on timescales and timekeeping, on clock comparisons, including optical-to-microwave comparisons with combs, and their applications. Several other aspects of our research are covered by other articles of the Special Issue of the Comptes Rendus de l'Acad\'emie des Sciences (Volume 16, Issue 5), i.e. the space mission PHARAO/ACES \cite{cras2014onACES}, fundamental tests with clocks \cite{cras2014onFundamentalTests}, development of technologies for space optical clocks \cite{cras2014onSpaceOpticalClocks} and optical fibers links \cite{cras2014onFiberLinks1,cras2014onFiberLinks2}.


\section{Atomic fountains} \label{fountains}

\begin{figure*}[t]
	\centering
  \includegraphics[width=1\textwidth, angle=0]{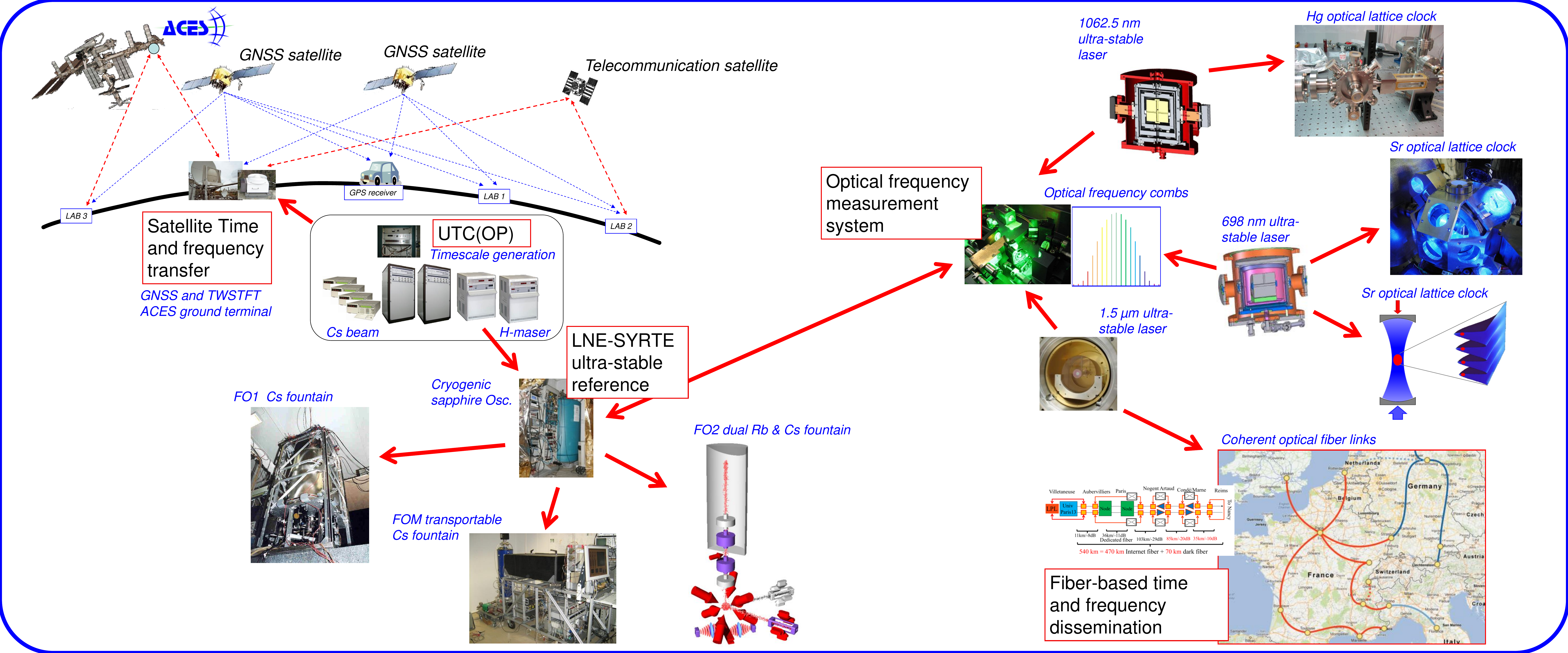}
	\caption{Overview of the LNE-SYRTE atomic clock ensemble at the Observatoire de Paris.}
	\label{fig:ClockEnsemble}
\end{figure*}

Atomic fountains are the first generation of the laser-cooled atomic frequency standards. They use the fountain geometry where spectroscopy of the clock transition is performed onto a free-falling sample of laser-cooled atoms which is beforehand launched upwards vertically (see, for instance, \cite{guena2012} and references therein). To date, atomic fountains using cesium provide the most accurate realization of the SI second. One important aspect of the last 10 years was to better understand systematic shifts limiting the accuracy of these devices.

\textit{Distributed cavity phase shift}-- In atomic fountains, the Ramsey interrogation is realized by the up-going and down-going passages of the atomic sample through the microwave Ramsey cavity. The spatial phase variations of the field inside the cavity, when sampled by the moving atoms, induce a frequency shift, which can be described as a residual Doppler shift. For a long time, there was a lack of both a complete and agreed model for this effect and of experiments to test it. Consequently, this effect was one of the main sources of uncertainty in atomic fountains. In \cite{li2004,li2010}, a new approach was proposed to compute the cavity phase distribution. We performed measurements of these shifts in FO2-Cs which enabled the first quantitative comparison between theory and experiment \cite{guena2011}. This study validated the theoretical model and lowered the distributed cavity phase uncertainty for FO2-Cs to $10^{-16}$. It also defined a method to determine this uncertainty, which was then adopted in \cite{li2011,weyers2012} and for other SYRTE fountains.

\textit{Microwave lensing shift}-- The microwave field inside the Ramsey cavity not only excites the transition between the two internal clock states but also modifies the motion of atomic wave packets, leading to a frequency shift \cite{Borde2002}\cite{Wolf2004}. In \cite{Gibble2006}, a new approach to compute the shift was proposed. It was then used in a complete model of the effect, taking into account all features of the interaction such as atomic velocity and space distributions and detection non-uniformities \cite{li2011}\cite{weyers2012}. This same method was applied to LNE-SYRTE fountains, for which shifts are reported in table 2 of \cite{guena2012}.

\textit{Blackbody radiation shift}-- In 2004, conflicting measurements and calculations of this shift induced by thermal radiation bathing the atoms were reported. This led us to revisit our early accurate measurements of the Stark coefficient \cite{simon1998}. Our new measurements at lower electric fields have been found to be in excellent agreement \cite{rosenbusch2007}. The theory of the Stark shift developed in the 60's turned out to have a sign error for the tensor part. This led to a small change of the blackbody radiation shift correction of $7\times 10^{-17}$ \cite{guena2012}. Two independent high accuracy ab initio calculations further agreed with the blackbody radiation shift correction derived from our Stark measurements.

\textit{Microwave leakage and synchronous phase perturbations}-- Interaction of atoms with unintended residual microwave field and synchronous perturbation of phase of the probing field can produce shifts. We developed microwave synthesizers that can be switched without introducing phase transients and a phase transient analyzer with 1~$\mu$rad.s$^{-1}$ resolution \cite{santarelli2009}. Using these tools, we lowered the uncertainty related to these putative frequency shifts to less than $10^{-16}$.

Our approach to deal with other systematic shifts remained as described in our last report in the Comptes Rendus \cite{Bize2004}. Table \ref{tab_accuracy} gives, as an example, the accuracy budget of FO2-Cs as of 2014.

\begin{table}
\footnotesize
\caption{Typical uncertainty budget of the FO2-Cs primary frequency standard (top). Uncertainty budget of the Sr1 optical lattice clock as of July 2011 (bottom). Tables give the fractional frequency correction and its Type~B uncertainty for each systematic shift, in units of $10^{-16}$. The total uncertainty is the quadratic sum of all uncertainties.}\label{tab_accuracy}
\begin{center}
\begin{tabular}{lcc}
\multicolumn{3}{c}{FO2-Cs}\\
\hline
Physical origin of the shift                                       & Correction   & Uncertainty  \\
\hline 
{\small Quadratic Zeeman}               & $-1919.9$&$0.3$  \\
{\small Blackbody radiation}                  &   $168.4$&$0.6$\\
{\small Collisions and cavity pulling}        &   $201.2$&$1.5$\\
{\small Distributed cavity phase}       & $-0.9$& $1.2$   \\
{\small Microwave lensing}                     &  $-0.7$&$0.7$   \\
{\small Spectral purity \& leakage}            & 0  &$0.5$ \\
{\small Ramsey \& Rabi pulling}                & 0 &$0.1$  \\
{\small Relativistic effects}            & 0 & $0.05$ \\
{\small Background collisions}                 & 0  &$1.0$ \\
\hline 
{\small Total}                               &  $-1551.9$&$2.5$  \\
\hline 
\end{tabular}
\hspace{1cm}
\begin{tabular}{lcc}
\multicolumn{3}{c}{Sr1}\\
\hline
Physical origin of the shift                                       & Correction   & Uncertainty  \\
\hline 
{\small Quadratic Zeeman}               & $19.7$&$0.2$  \\
{\small Blackbody radiation}                  &   $ 53.8 $&$0.8$\\
{\small Collisions}        &   $-0.2 $&$0.5 $\\
{\small AC Stark shift lattice 1st order}       & $-0.5 $& $0.1 $   \\
{\small AC Stark shift Lattice 2nd order}      &  $0 $&$ 0.1$   \\
{\small DC Stark shift}            &  0  &$0.01 $ \\
{\small Line pulling}                & 0 &$0.5$  \\
\hline 
{\small Total}                               &  $72.8$&$1.0$  \\
\hline 
\end{tabular}
\end{center}
\vspace{-6mm}
\end{table}

\normalsize
\vspace{1mm}

Another important achievement was the simultaneous operation with $^{87}$Rb and $^{133}$Cs of the dual fountain FO2. This was done by implementing dichroic collimators overlapping 780~nm (for Rb) and 852~nm (for Cs) radiations for all laser beams, and by adopting a time sequence enabling time resolved selective detection of the two atomic species \cite{guena2010}. Further notable improvement relates to reliability and capability for long term unattended operation. Using 2D magneto-optic traps to load the optical molasses suppressed residual background vapor and enhanced the lifetime of the alkali sources. Also, we developed an automatic data processing system that monitors the status of all fountains, oscillators and internal links in quasi real time. This system allows rapid detection of failures. It also performs automated fountain data processing, taking account of all systematic corrections, and it continuously generates frequency measurements at the nominal uncertainty of the fountains. This capability has had major impact on timekeeping and other applications of atomic fountains (see sect.~\ref{sec:fundphystests} and \ref{sec:timekeeping} ).
%
%
%
%
%

\section{Optical lattice clocks}

\label{sec:OLC}

In optical lattice clocks (OLCs), a set of neutral cold atoms, dipole-trapped in an optical lattice, are interrogated by an ultra-stable ``clock'' laser. Because they involve probing an optical transition of a large (typically $10^4$) number of tightly confined atoms, they combine an excellent ultimate frequency stability -- only limited by the Quantum Projection Noise (QPN) -- and a high accuracy. Proposed in 2001 \cite{Katori2001}\cite{Katori2003a}, OLCs made tremendous progress in the last decade. OLCs have demonstrated unprecedented frequency stabilities of a few $10^{-16}/\sqrt{\tau}$ and a record accuracy below $10^{-17}$ \cite{bloom2014optical}, overcoming the best ion clocks~\cite{letargat2013,hinkley_atomic_2013,falke2014,bloom2014optical,ushijima2014cryogenic}. With current improvement in laser stabilization, OLCs are expected to reach a QPN limited stability on the order of $10^{-17}/\sqrt{\tau}$ within a few years, thus enabling even better characterization of systematic effects. OLCs with Sr, Yb, Hg and more prospectively Mg have been demonstrated. Among these atomic species, Sr is currently the most popular choice because of the accessibility of the required laser wavelengths, the possibility to cool Sr down to sub-$\mu$K temperature using the narrow $^1S_0 \rightarrow{}^3P_1$ inter-combination line, and the possibility it offers on the control of systematic effects, most notably concerning  the high order perturbation by the trapping light and cold collisions.

\textit{Strontium optical lattice clocks}-- LNE-SYRTE developed two OLCs using strontium atoms. The design of these clocks uses an optical cavity to enhance the optical lattice light, giving access to large trap depths. It enabled us to explore systematic effects induced by the trapping laser light. These effects are specific to OLCs, and we have demonstrated that they can be controlled to better than $10^{-17}$, even with a significant trapping depth~\cite{Brusch2006a,westergaard2011lattice}, thus validating the concept of OLCs. In particular, we have determined the precise value for the ``magic wavelength'' for which the impact of the trapping light is canceled to first order, and resolved or upper-bounded a number of higher order effects. Because OLCs use a large number of atoms in a tightly confined space, they are subject to a significant density-dependent systematic frequency shift. Some groups have resolved a density shift on the order of $10^{-16}$ with both Sr and Yb. However, the loading technique chosen at LNE-SYRTE leads to a lower atomic density, thus dramatically reducing this effect below $10^{-17}$. The blackbody radiation shift has remained the dominant contribution to the accuracy budget, with an uncertainty around $5\times 10^{-17}$ for both Sr and Yb, assuming a 1~K uncertainty on the temperature of the environment. Recently, precise measurements of the static polarizability of Yb and Sr, together with carefully crafted environments for the atoms has enabled a few groups to drastically reduce this uncertainty, down to the $10^{-18}$ range.

Comparisons between clocks are necessary to confirm their accuracy budget. The first comparisons between remote Sr OLCs were achieved by comparing to cesium clocks (see sect.~\ref{sec:fundphystests}), but they were soon limited by the cesium accuracy. LNE-SYRTE published the first comparison between two local OLCs that confirm the accuracy budget of the clocks better than the accuracy of the cesium fountains, involving two Sr clocks with a frequency difference smaller than their combined accuracy budgets of $1.5\times 10^{-16}$~\cite{letargat2013}. This resolution is obtained after less than one hour of integration. Table~\ref{tab_accuracy} gives the accuracy budget of one of the Sr OLCs at the time of this comparison.


\textit{Mercury optical lattice clock}-- SYRTE also started the development of a mercury OLC. Hg has the advantage of low sensitivity to blackbody radiation and to electric field (30 times lower than Sr, 15 times lower than Yb). For the $^{199}$Hg isotope, clock levels have a spin 1/2 for which the tensor light shift sensitivity is absent. Also, because of its high vapor pressure, it does not require an oven and enables the use of a 2D magneto-optic trap as the initial source of atoms. Hg is also interesting for fundamental physics and atomic physics, because of a quite high sensitivity to a variation of $\alpha$ (see sect.~\ref{sec:fundphystests}) and its 7 natural isotopes. The main challenge of using Hg lies in the need for deep UV laser sources. When the potential of Hg for a highly accurate optical lattice clock arose, Hg had never been laser cooled.

In the last years, we made all the steps leading to the demonstration, for the first time, of a Hg lattice clock. Laser cooling on the 254 nm $^1S_ 0 \rightarrow{} ^3P_1$ intercombination transition was demonstrated and studied \cite{Petersen2008b}\cite{Hachisu2008}\cite{McFerran2010}. A clock laser system with thermal noise limited instability of $4\times 10^{-16}$ was developed \cite{Millo2009b}\cite{Dawkins2010}. We performed the first direct laser spectroscopy of the the clock transition, firstly on atoms free-falling from a magneto-optic trap \cite{Petersen2008a} and secondly on lattice-bound atoms, with linewidth down to 11~Hz (at 265.6~nm or 1128~THz). We performed the first experimental determination of the ``magic wavelength'' \cite{Yi2011}, for which our best value is $362.5697 \pm �0.0011$~nm. We performed initial measurements of the absolute frequency of the $^{199}$Hg clock transition down to an uncertainty of 5.7 parts in $10^{15}$ \cite{McFerran2012}.

So far, the advancement of the Hg optical lattice was mainly hindered by the poor reliability of the 254~nm laser-cooling and by the modest lattice trap depth. Recent work enabled large improvements to overcome these two limitations, opening the way to in-depth systematic studies and higher accuracies.

\section{Optical frequency combs}
{}
%
%
The development of optical frequency standards of the kind presented in the previous section is aimed at producing a laser electromagnetic field of extremely stable and accurate frequency. To realize a complete metrological chain that allows comparison with other standards operating at different wavelengths in the optical domain or in the microwave domain (such as primary frequency standards), it is necessary to use a specific device. The method of choice nowadays is the optical frequency comb based on mode-locked femto-second laser, that provides a phase coherent link spanning across the optical and microwave domains. Recently, SYRTE focused on a technology of comb based on erbium-doped fiber lasers, whose foremost asset is the capability to function for months with a very limited maintenance while performing state-of-the-art measurements. Tight phase locking (bandwidth $\sim 1\,$MHz) of such comb onto a 1.5~$\mu$m ultra-stable laser sets it in the narrow-linewidth regime where beatnotes with other ultra-stable light at other wavelengths are typically $\sim1\,$Hz linewidth. This provides high performance simultaneous measurements of the various ultra-stable optical frequency references at SYRTE in the visible and near infrared domain.

The comb also behaves like a frequency divider: the repetition rate of the comb, $f_{\rm rep}$, results from the coherent division of the 1.5$\,\mu$m laser frequency $\nu_{\rm L}$. By photo-detecting the train of pulses and filtering out one specific harmonic of the repetition rate, one can generate a microwave signal whose phase noise is that of the cw optical reference, divided by the large frequency ratio (typically 20000) between a 1.5$\,\mu$m wavelength cw laser and a $10\,$GHz microwave signal. $10\,$GHz signals with phase noise of -100\,dBc/Hz at 1\,Hz Fourier frequency and -140\,dBc/Hz white noise plateau are now straightforward to produce. This low phase noise level is comparable with that of cryogenic sapphire oscillators and is thus sufficient to operate state-of-the-art microwave atomic fountains (sect.~\ref{fountains}) with a short term stability limited only by atomic quantum projection noise. We have realized proof-of-principle experiments of such a scheme \cite{MilloAPL2009} and are now progressing toward implementing it in an operational system. We further demonstrated several advanced techniques \cite{ZhangAPB2012, HabouchaOL2011, ZhangOL2014} to reduce the imperfection of the frequency division and photo-detection processes. We have shown that it is now becoming possible to generate microwave signals with phase noise as low or lower than any other technology for a large range of Fourier frequencies. Applications of such extremely low noise microwave signals can be found in RADAR (civil and military) as well as very long baseline interferometry.

\begin{figure*}[ht]
	\centering
  \includegraphics[width=0.9\textwidth, angle=0]{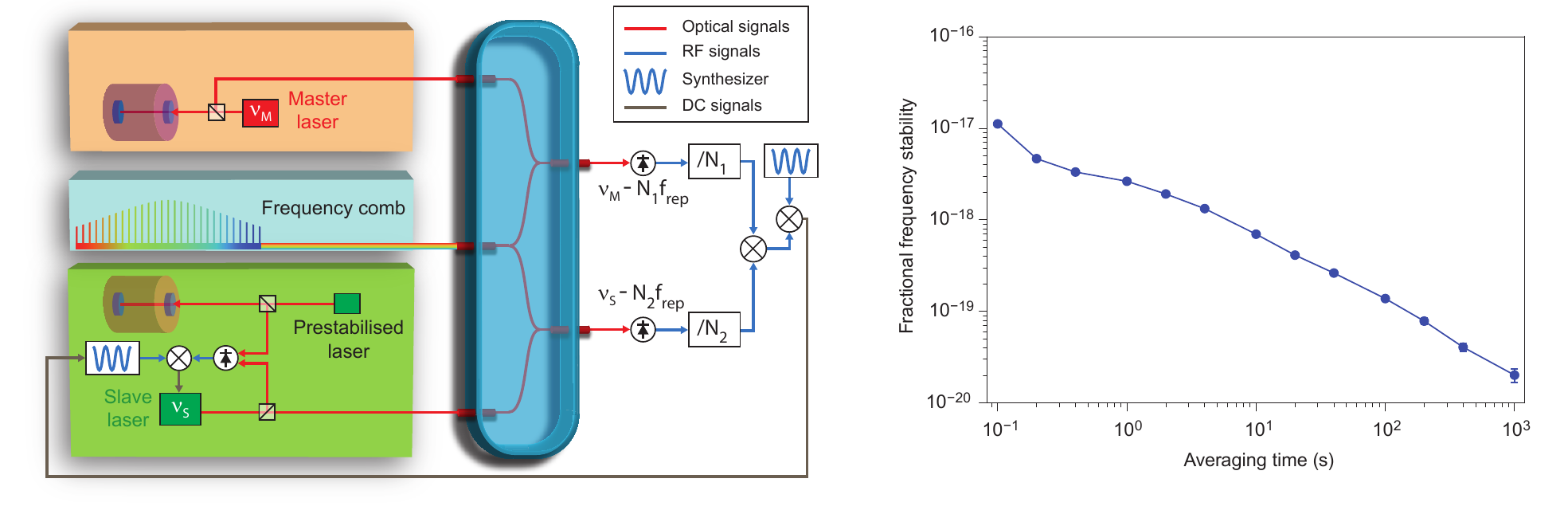}
	\caption{Principle of the transfer of spectral purity (left): the optical beatnotes of the comb with a master laser on the one hand, and with a slave laser on the other hand, are rescaled and mixed before being compared to a stable synthesizer. The feedback on the offset-locked slave transfers the spectral purity of the master to the slave laser. The modified Allan deviation (right) of the noise added by the transfer itself has a level of only $2\times10^{-18}$ at $1\,$s, and averages down to $2 \times 10^{-20}$ after $1000\,$s.}
	\label{fig:TSP}
\end{figure*}

Finally, the comb-based transfer of spectral purity between different wavelengths recently led to exciting results. This application is crucial for the future development of optical lattice clocks, whose short term stability is currently limited by the spectral purity of the ultra-stable clock laser probing the atomic transition. Several competing technologies are being explored, notably at SYRTE, to improve the performance of these lasers, all of them very challenging, some of them wavelength-specific. Being able to utilize the performance of an extremely stable laser at a given wavelength and to distribute its performance to any wavelength within reach of the frequency comb is an important milestone for the future developments of optical lattice clocks. We demonstrated such transfer from a master to a slave laser with an added instability of no more than a few $10^{-18}$ at $1\,$s (see fig. \ref{fig:TSP}, right), well within the requirements expected for the next several years \cite{NicolodiNP2014}.

\section{Fundamental physics tests}\label{sec:fundphystests}

One exciting scientific application of atomic clocks with extreme uncertainties is to contribute to testing fundamental physical laws and searching for physics beyond the Standard Model of particle physics. The frequency of an atomic transition relates to parameters of fundamental interactions (strong interaction, electro-weak interaction), such as the fine-structure constant $\alpha$, and to fundamental properties of particles like for instance the electron mass, $m_e$. Repeated highly accurate atomic clock comparisons can be used to look for a putative variation with time or with gravitational potential of atomic frequency ratios, and, via suitable atomic structure calculations, of natural constants. Clocks provide laboratory tests, independent of any cosmological model, that constrain alternative theories of gravity and quantum mechanics, thereby contributing to the quest for a unified theory of the three fundamental interactions.

\begin{figure}
	\begin{center}

	 \includegraphics[width=0.45\textwidth]{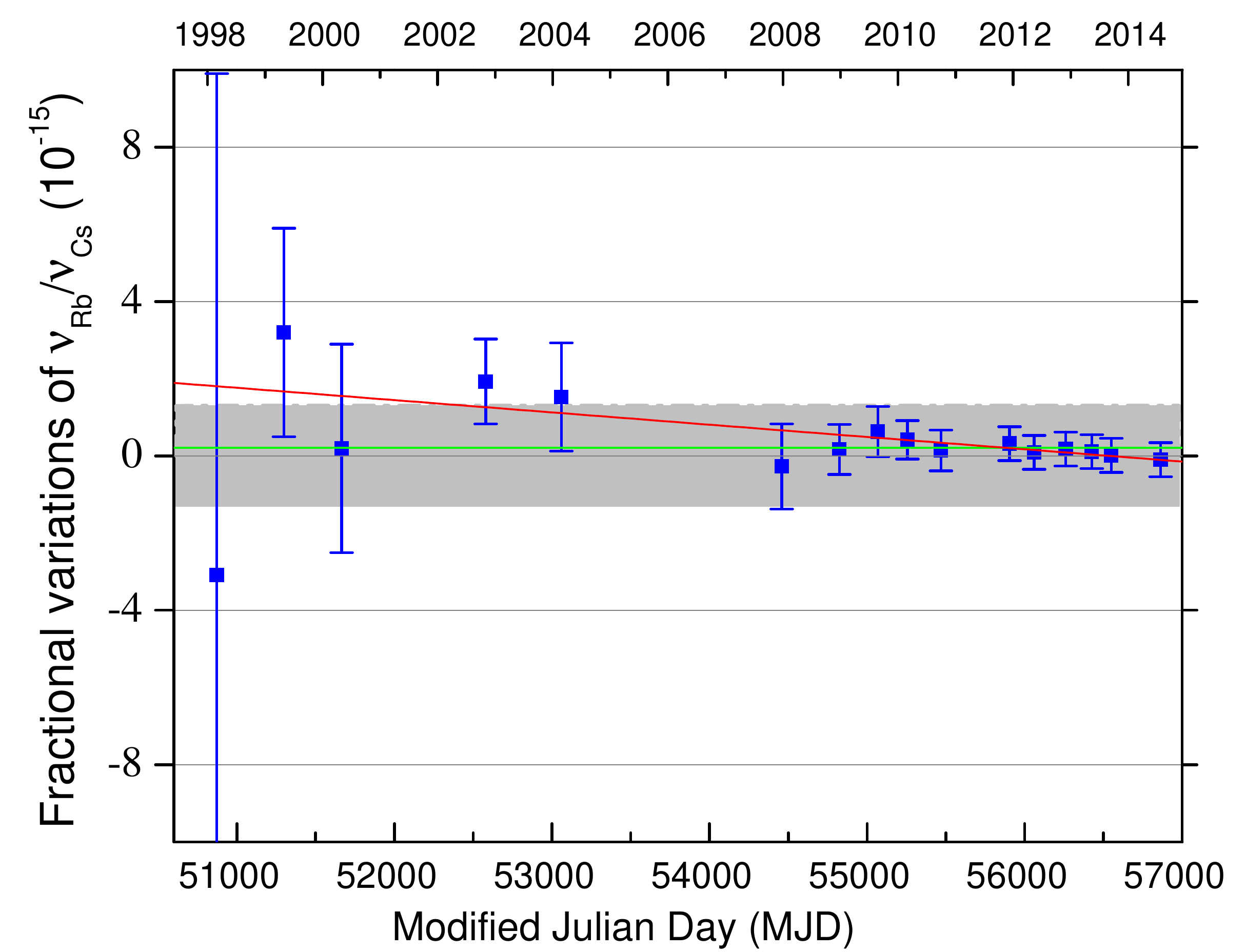} 

	 \vspace{3mm}

 \includegraphics[width=0.48\textwidth]{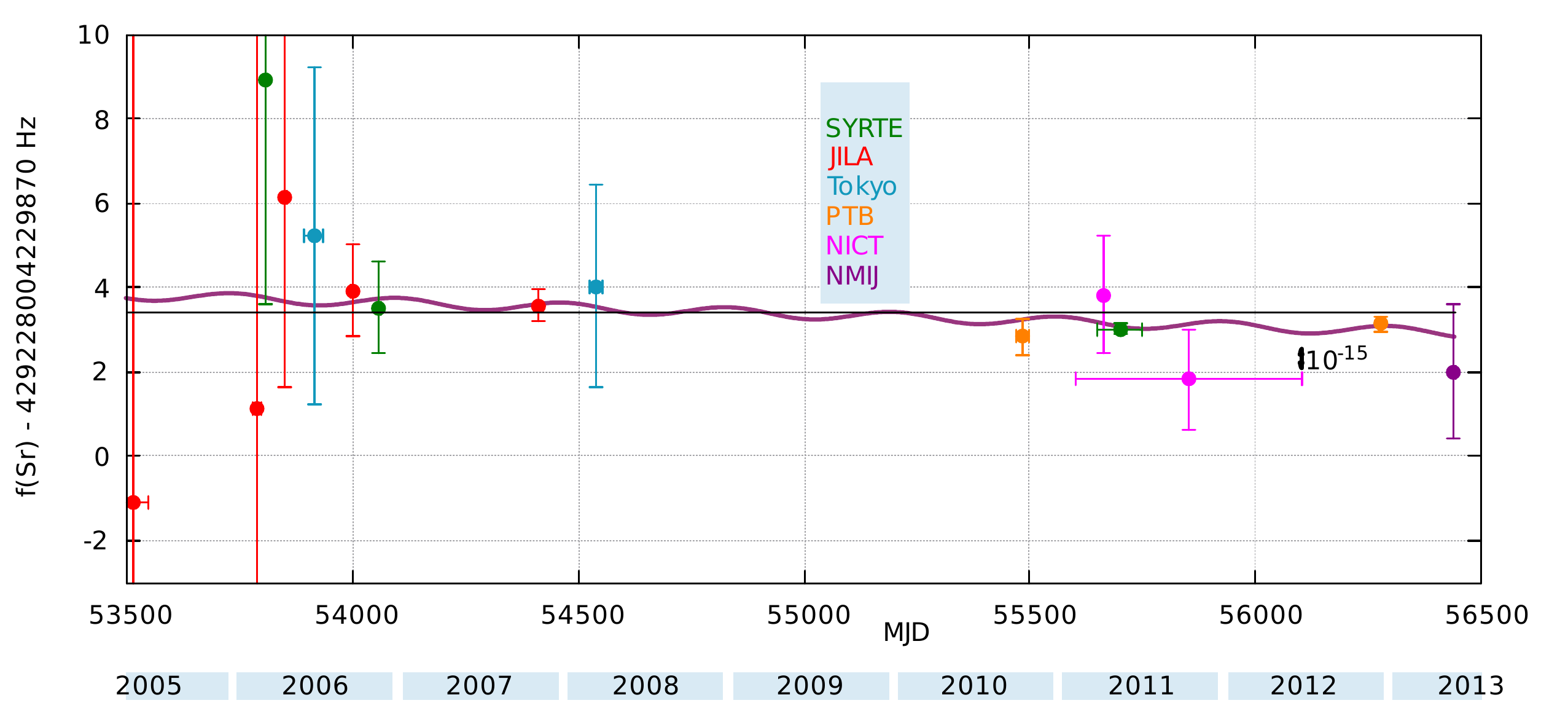}
 
	\end{center}

	 \vspace{1mm}

\footnotesize
\begin{center}
\begin{tabular}{p{28mm} ccc}
\hline \hline
 & $\ln(\alpha)$ &$\ln(\mu$)&$\ln(m_{q}/\Lambda_{\mathrm{QCD}})$ \\
\hline
$d/dt~ (\times 10^{-16}\mathrm{yr}^{-1})$& $-0.26\pm 0.24$ &$1.1\pm 1.4$& $59\pm 30$\\
$c^{2}d/dU ~(\times 10^{-6})$ & $0.27 \pm 0.46$ & $-0.2 \pm 2.1$ & $-2.9\pm 5.6$ \\
\hline
\end{tabular}
\end{center}
\normalsize

\caption{Top: Temporal record of fractional variations of the $\nu_{\mathrm{Rb}}/\nu_{\mathrm{Cs}}$ hyperfine frequency ratio. The error bars are the total 1~$\sigma$ uncertainties, dominated by the systematic uncertainties. The solid red line is the weighted fit to a line with inverse quadratic weighting. The origin of the vertical axis corresponds to the $^{87}$Rb secondary representation of the SI second recommended in 2012, with a recommended uncertainty $1.2 \times 10^{-15}$ (grey area) \cite{CCTF2012}. Middle: International comparisons of the $\nu_{\mathrm{Sr}}/\nu_{\mathrm{Cs}}$ frequency ratio. A fit shows an upper bound on a drift of this ratio, as well as on a variation synchronized with the Earth's orbit around the Sun. Bottom: Results of the global analysis of accurate experimental determinations of variations of atomic frequency ratios, available as of October 2014. The table gives constraints on temporal variations and on couplings to gravitational potential for the three fundamental constants: $\alpha$, $\mu=m_{e}/m_{p}$ and $m_{q}/\Lambda_{\mathrm{QCD}}$.} \label{Graphdrift}
\end{figure}

\label{RbCsFitLin}
\textit{$^{87}$Rb vs $^{133}$Cs comparisons}-- Improvements of atomic fountains described in sect.~\ref{fountains} enabled major enhancement in the number and in the quality of Rb/Cs hyperfine frequency ratio measurements since our last report in the Comptes Rendus \cite{Bize2004}. Measurements have been performed almost continuously since 2009. Fig.\ref{Graphdrift}, top shows the temporal record of the variations of this ratio. Measurements extending over 14~years give stringent measurements of a putative variation with time and gravity
of the Rb/Cs ratio, as reported in \cite{guena2012b}. Taking into account out most recent data, we get $d \ln(\nu_{\mathrm{Rb}}/\nu_{\mathrm{Cs}})/dt=(-11.6\pm 6.1 )\times 10^{-17}$~yr$^{-1}$ for the time variation. For the variation scaled to the annual change of the Sun gravitational potential on Earth $U$, we get $c^2 d \ln(\nu_{\mathrm{Rb}}/\nu_{\mathrm{Cs}})/dU=(7.4 \pm 6.5)\times10^{-7}$, which provides a differential redshift test between Rb and Cs twice more stringent than \cite{peil2013}.

\textit{$^{87}$Sr vs $^{133}$Cs comparisons}-- Frequency ratios between optical and microwave clocks offer a different sensitivity to natural constants than hyperfine frequency ratios. International absolute frequency measurements of strontium optical lattice clocks against Cs fountain primary frequency standards over a decade (fig.\ref{Graphdrift}, Middle) give the linear drift with time of the $\nu_{\mathrm{Sr}}/\nu_{\mathrm{Cs}}$ ratio: $d \ln(\nu_{\mathrm{Sr}}/\nu_{\mathrm{Cs}})/dt=(-2.3 \pm 1.8)\times 10^{-16}$~yr$^{-1}$, and of the variation with the gravitational potential: $c^2 d \ln(\nu_{\mathrm{Sr}}/\nu_{\mathrm{Cs}})/dU=(-1.3 \pm 1.5)\times 10^{-6}$. Because the accuracy of these measurements has improved considerably over time, these bounds will be significantly improved by future measurements.

\textit{Combining with other comparisons}-- Each pair of atoms has a different sensitivity to variations of three fundamental constants $\alpha$, $\mu=m_e/m_p$ and $m_{q}/\Lambda_{\mathrm{QCD}}$. To set independent limits to variations of these three constants with time, one can perform weighted least-squares fit to all accurate experimental determinations of variation with time of atomic frequency ratios, available as of October 2014. This includes the above Rb/Cs result (see sect.~\ref{RbCsFitLin}), optical frequency measurements of H(1S-2S) \cite{fischer2004}, Yb$^{+}$ \cite{tamm2014}, Hg$^{+}$ \cite{fortier2007}, Dy \cite{leefer2013} and Sr (see above, and \cite{letargat2013} and references therein) against Cs, and the optical-to-optical ion clock frequency ratio $\mathrm{Al}^{+}/\mathrm{Hg}^{+}$ \cite{rosenband2008}.
The fit yields independent constraints for the three constants given in the first row of the Table in fig.~\ref{Graphdrift}. The constraint relative to $\alpha$ is mainly determined by the $\mathrm{Al}^{+}/\mathrm{Hg}^{+}$ comparison. In this fit, only the Rb/Cs comparison disentangles variations of $\mu$ and of $m_q/\Lambda_{\mathrm{QCD}}$. It is therefore essential to constrain $m_q/\Lambda_{\mathrm{QCD}}$. This stems from the fact that optical frequency measurements are all performed against the Cs hyperfine frequency, except $\mathrm{Al}^{+}/\mathrm{Hg}^{+}$.

Similarly, we perform a global analysis for the variation with the gravitational potential exploiting all available comparisons as of October 2014 \cite{peil2013, fortier2007, leefer2013} and the above Rb/Cs and Sr/Cs results.
The least-squares fit to these results yields independent constraints for the three couplings to gravity given in the second row of the Table in fig.~ \ref{Graphdrift}.

The number of atomic systems contributing to improve these tests will continue to grow, e.g. with $^{88}$Sr$^+$ \cite{Madej2012}\cite{Barwood2014} and $^{171}$Yb \cite{Lemke2009a}, thanks to the steady efforts of many laboratories worldwide in the field of optical frequency metrology.


\section{Advanced timekeeping}\label{sec:timekeeping}

\begin{figure}[htb]

\includegraphics[width=0.45\textwidth]{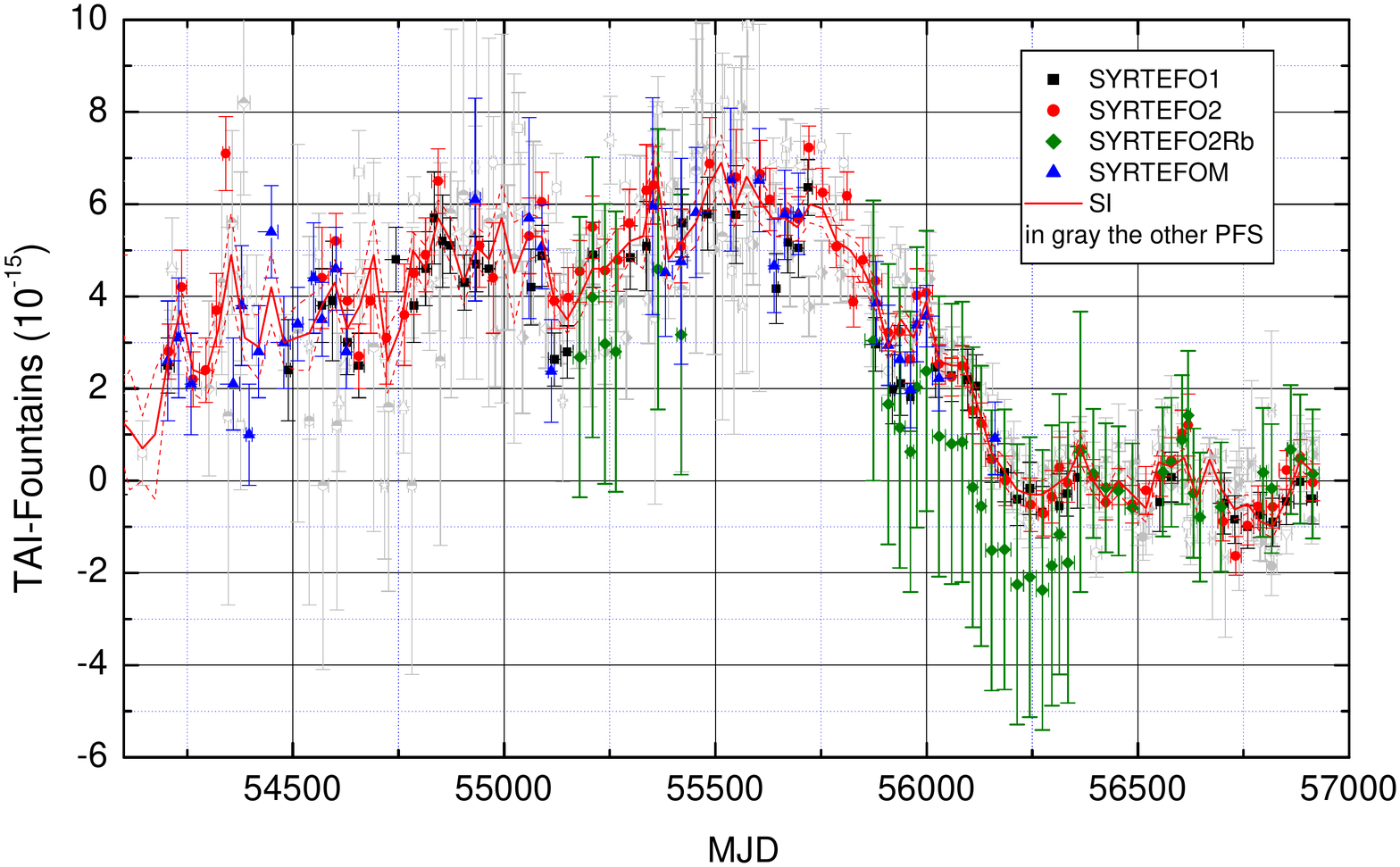}
\includegraphics[width=0.45\textwidth]{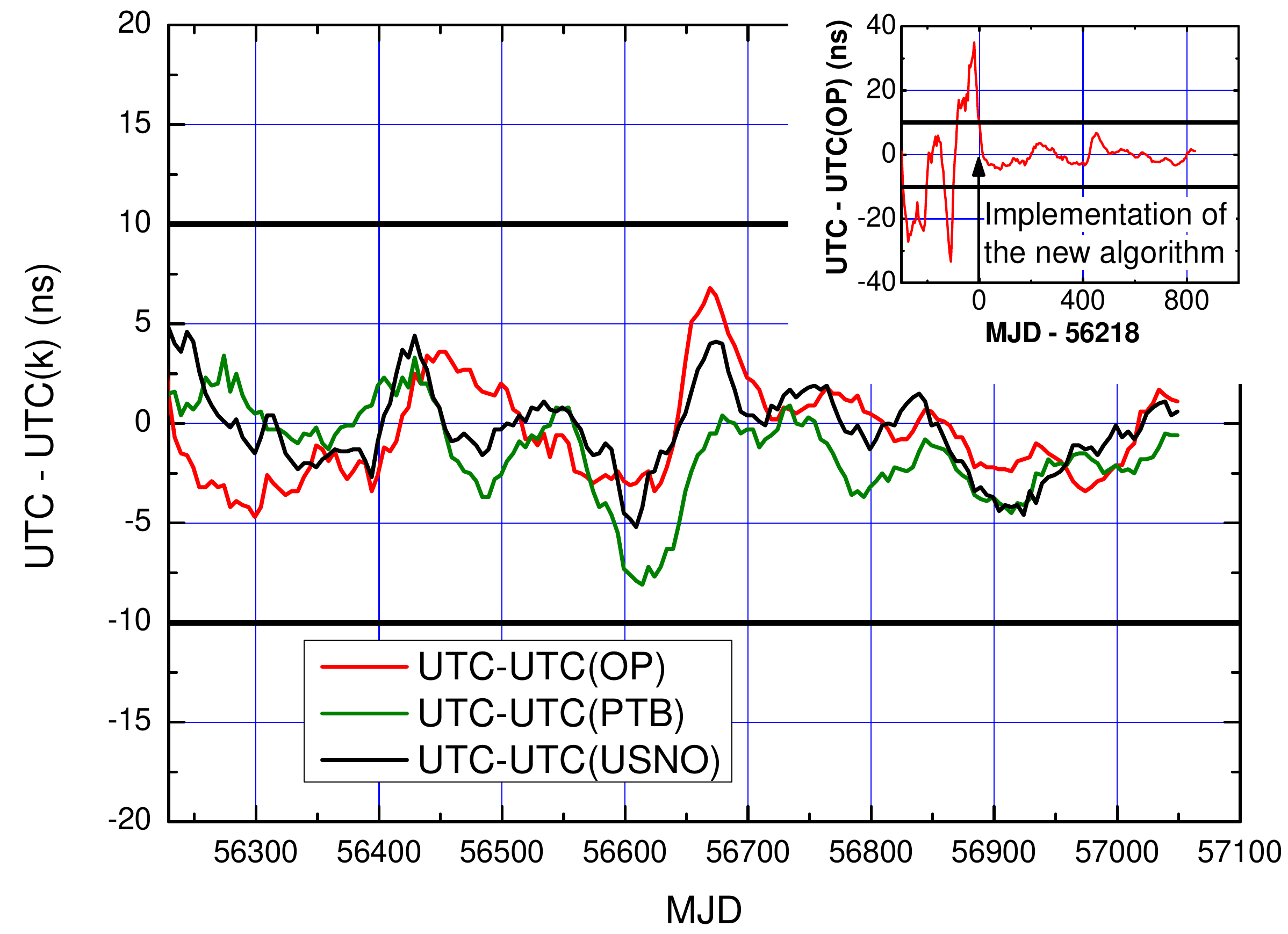}
\caption{Top: Calibrations of TAI by the atomic fountain PFSs. Filled symbols: contributions of SYRTE fountains. Solid red line: the SI.  Bottom: Comparisons of 3 UTC(k) to UTC: UTC(OP), UTC(PTB) and UTC(USNO). The inset shows the significant improvement achieved with the new method for generating UTC(OP) implemented at MJD 56218, compared to the previous one using a commercial Cs clock manually steered towards UTC.}
\label{TAI_UTC}
\end{figure}

\textit{TAI calibration with atomic fountains}-- The International Atomic Time (TAI) which is based on approximately 400 atomic clocks, now gets its accuracy from some ten atomic fountain clocks worldwide (see e.g. \cite{cras2014Petit}). In the last decade, the number of calibrations of TAI with atomic fountains has grown from approximately 10 per year to 4 to 6 per month in 2014, while simultaneously the accuracy improved from several $10^{-15}$ to a few $10^{-16}$, improving TAI a lot. Combining a tremendous number of monthly calibrations and a high accuracy, LNE-SYRTE atomic fountains are providing the largest contribution to the accuracy of TAI. Between 2007 and August 2014, they provided 197 calibrations out of a total of 407 calibrations worldwide, a weight of nearly 50\%. fig.~\ref{TAI_UTC}, Top shows these calibrations as published in \emph{Circular T}, and the SI second (red line) which is the average over all primary calibrations computed by the BIPM on a monthly basis. This illustrates how research on laser-cooled atomic fountain started 25 years ago led to improving an important service and infrastructure for science and society.

%
%

\textit{UTC(OP): timescales using atomic fountain clocks}-- The UTC(OP) timescale, elaborated at SYRTE, in Observatoire de Paris, is the real time realization of UTC for France. It is a continuously operated time reference used for multiple purposes:
definition of legal time disseminated in France, reference provided to French laboratories for
synchronization applications, pivot for French contributions to TAI, test of advanced time transfer
methods, link to UTC of the EGNOS system, contribution to the development of GALILEO, time reference for the ground-segment of the PHARAO/ACES space mission \cite{laurent2006, cacciapuoti2007,cacciapuoti2009}.

Progress in the accuracy and most importantly in the reliability of SYRTE atomic fountains (see sect.~\ref{fountains}) enabled a new implementation of UTC(OP). A new UTC(OP) algorithm based on a hydrogen maser steered by the atomic fountains was developed and
implemented in October 2012. The maser is predictable enough to reach a stability of $\sim 10^{-15}$. The
atomic fountains allow the maser frequency to be calibrated with an uncertainty in the $10^{-16}$ range \cite{guena2012, guena2014}. These features are sufficient to maintain a phase deviation of a few ns over $1-
2$ months, which corresponds to the delay of publication of the BIPM \emph{Circular T}.
This timescale is as autonomous and independent as possible, except a small long term steering to remain
close to UTC, and does not rely on any other timescale available in real time such as GPS time or
other UTC(k). A timescale with these characteristics provides a powerful tool to understand current limits and eventually improve international timekeeping.

Practically, UTC(OP) is realized using a microphase stepper fed by the reference maser. A frequency
correction is updated every day to compensate the maser frequency and
maintain UTC(OP) close to UTC. This correction is the sum of two terms. The main term corresponds to the current frequency of the maser as measured
by the fountains. The value is estimated with a linear extrapolation of the data covering the past 20~days
to remain robust against possible interruptions of data provision or of the automatic data processing. The
second term is a fine steering to maintain UTC(OP) close to UTC, compensating the
frequency and phase offset between UTC(OP) and UTC. It is updated monthly at the BIPM \emph{Circular~T}
publication. The steering correction is usually of the order of $10^{-15}$ or below.

Figure \ref{TAI_UTC}, bottom presents the comparison of three UTC(k) to UTC as published in \emph{Circular~T}
since the implementation of the new UTC(OP). Over this period, UTC(OP) is one of the 3 best real time
realizations of UTC \cite{rovera2013, abgrall2014}, with UTC(PTB), the pivot of time
transfers for international contributions to TAI, and UTC(USNO), the laboratory providing the largest number of clock data included in EAL computation. Departure between UTC(OP) and UTC remains well below 10~ns, with a rms value less than 3~ns. This is an improvement of about a factor of 5 compared to the previous realization method of the timescale. On-going instrumental upgrades shall further improve the short term stability of the timescale and the robustness of the system.


\section{Toward a redefinition of the SI second}
{}

Several optical frequency standards are now largely surpassing Cs atomic fountains which realize the second of the international system of units (SI) and define the accuracy of TAI. This opens the inviting prospect of a redefinition the SI second. In 2001, anticipating this situation, the Consultative Committee for Time and Frequency (CCTF) of the Comit\'e International des Poids et Mesures (CIPM) recommended that a list of Secondary Representations of the Second be established. Secondary Representations of the SI Second (SRS) are transitions which are used to realize frequency standards with excellent uncertainties, and which are measured in the SI system with accuracies close to the limit of Cs fountains. They are part of the broader list of recommended values of standard frequencies produced and maintained by the CCL-CCTF Working Group and adopted by the CIPM. Producing and maintaining these lists of recommended values is a vehicle to keep track of measurements providing the most stringent connections between the optical domain and the SI second, and to verify the level of consistency between these measurements. This is an important task to prepare for a possible redefinition of the SI second.


\textit{Contributions to the list of recommended values}--
LNE-SYRTE provided several high accuracy absolute frequency measurements which contributed to the list of recommended values. The $^{87}$Rb hyperfine transition was measured repeatedly against Cs fountains, as already shown in fig.~\ref{Graphdrift}. This transition became the first Secondary Representation of the Second proposed by the CCTF in 2004, based our early measurements, and adopted by the CIPM in 2006. After further significant progress visible in fig.~\ref{Graphdrift}, the recommended value was revised by the 2012 CCTF and adopted in 2013. The origin of the vertical scale of the graph is the recommended value of 2012 and the gray area represents the recommended uncertainty. The weighted average of all data points (green line) gives $(2.15\pm 1.48)\times 10^{-16}$ is consistent with zero within the smallest overall uncertainty of the measurements ($4.4\times 10^{-16}$). LNE-SYRTE also provided absolute frequency measurements that contributed to the establishment of the recommended values for the $^1S_0\rightarrow{}^3P_0$ of $^{87}$Sr \cite{Baillard2007b}, $^{88}$Sr \cite{Baillard2007} and $^{199}$Hg \cite{McFerran2012}. Using the transportable Cs fountain primary standard FOM, LNE-SYRTE also contributed to the establishment of the recommended value for H(1S-2S) (measured at the  Max Planck Institut f\"ur Quantenoptik, Garching, Germany) \cite{parthey2011} and $^{40}$Ca$^{+}$ (at the University of Innsbruck, Austria) \cite{chwalla2009}.

$^{87}$Sr is currently the most widespread optical frequency standard. For this reason a large number of groups measured the absolute frequency of $^{87}$Sr against Cs with a remarkable degree of consistency, as can be seen in fig.~\ref{Graphdrift}, middle. These measurements led the 2006 CCTF to recommend the $^{87}$Sr as a Secondary Representation of the SI second. The recommended value was updated by the 2012 CCTF based on measurements from 5 institutes. It has a recommended uncertainty of $1\times 10^{-15}$. More recently, LNE-SYRTE reported a measurement of the $^{87}$Sr clock transition with an uncertainty of $3.1\times 10^{-16}$ limited by the accuracy of atomic fountains \cite{letargat2013}. This is the most accurate absolute measurement to date of any atomic frequency. One of the key factors for the measurement is the record stability between an optical and a microwave clock: $4.1\times 10^{-14}/\sqrt{\tau}$ against Cs (and $2.8\times 10^{-14}/\sqrt{\tau}$ against Rb). In 2014, PTB reported another absolute frequency measurement with an uncertainty of $3.9\times 10^{-16}$ \cite{falke2014}. These two last measurements are in excellent agreement.




\textit{Using a Secondary Representation of Second to calibrate TAI}-- One major application of primary frequency standards is to calibrate and steer the scale interval of the widely used International Atomic Time TAI. It is important to anticipate how a possible redefinition of the second would impact the elaboration of TAI. We used the FO2-Rb fountain to investigate how a Secondary Representation of Second could participate to TAI. Calibrations of the frequency of our reference hydrogen maser were produced with FO2-Rb, in a similar way that absolute calibrations are done with primary frequency standards. These data were then submitted to the BIPM and to the Working Group on Primary Frequency Standards. Following this submission, the BIPM and the Working Group defined how frequency standards based on Secondary Representations will be handled by the BIPM and how they will be included into the \textit{Circular~T}. The Working Group was renamed Working Group on Primary and Secondary Frequency Standards and it was decided that calibrations produced by LNE-SYRTE with FO2-Rb could be included into \textit{Circular~T} and, since July 2013, contribute to steering TAI. This was the first time that a transition other than the Cs hyperfine transition was used to steer TAI \cite{guena2014}.

\textit{Absolute frequency measurement against the TAI ensemble}-- More than 40 formal calibrations of TAI with FO2-Rb have been sent, processed by the BIPM and published into \textit{Circular~T}. These data can be used to measure the frequency Rb hyperfine transition directly against the second as realized by the TAI ensemble. This can be done with a statistical uncertainty of 1 part in $10^{-16}$, and therefore at the accuracy limit of primary frequency standards defining the scale interval of TAI \cite{guena2014}. This illustrates how TAI provides worldwide access to the accuracy of Cs fountains. This also shows how recommended values of Secondary Representation of the Second based on optical transitions could be checked against the SI second as realized by the TAI ensemble.



\section{Prospects}

In the future, highly accurate atomic clocks and their applications will keep improving at a high pace. An important milestone in the field will be the simultaneous availability of advanced timescales, of the new generation of optical clocks and of the means to compared them remotely at unprecedented levels of uncertainty. In the coming decade, the ACES mission will allow ground-to-space comparisons to the $10^{-16}$ level and ground-to-ground comparisons to the mid $10^{-17}$ level \cite{cras2014onACES}. Optical fiber links will allow comparisons of the most accurate optical clocks at their limit: $10^{-18}$ or better \cite{cras2014onFiberLinks1,cras2014onFiberLinks2}. We can confidently predict major improvements in all applications of highly accurate atomic clocks. Availability of clocks and clock comparisons at the $10^{-18}$ level can further enable new applications. Clock comparisons determine Einstein's gravitational redshift between the 2 remote clock locations. For a clock at the surface of the Earth, $10^{-18}$ corresponds to an uncertainty of 1~cm in height-above-geoid, making the idea of clock-based geodesy realistic and potentially useful. Highly accurate clocks could become a new type of sensors for applications in Earth science, illustrating once again the fertilizing power of the quest for ever increased accuracy.

%
%
%

\section*{Acknowledgements}

SYRTE is UMR CNRS 8630 between Centre National de la Recherche Scientifique (CNRS), Universit\'e Pierre et Marie Curie (UPMC) and Observatoire de Paris. LNE, Laboratoire National de M\'etrologie et d'Essais, is the French National Metrology Institute.
SYRTE is a member of IFRAF, of the nanoK network of the R\'egion \^Ile de France and of the FIRST-TF LabeX. We acknowledge the large number of contributions of SYRTE technical services. This work is supported by LNE, CNRS, UPMC, Observatoire de Paris, IFRAF, nanoK, Ville de Paris, CNES, DGA, ERC AdOC, EMRP JRP SIB55 ITOC, EMRP JRP EXL01 QESOCAS. We are grateful to the University of Western Australia and to M.E.~Tobar for the long-lasting collaboration which gives us access to the cryogenic sapphire oscillator used in the LNE-SYRTE ultra-stable reference.

\bibliography{LNESYRTE2014}

\begin{thebibliography}{63}%
\makeatletter
\providecommand \@ifxundefined [1]{%
 \@ifx{#1\undefined}
}%
\providecommand \@ifnum [1]{%
 \ifnum #1\expandafter \@firstoftwo
 \else \expandafter \@secondoftwo
 \fi
}%
\providecommand \@ifx [1]{%
 \ifx #1\expandafter \@firstoftwo
 \else \expandafter \@secondoftwo
 \fi
}%
\providecommand \natexlab [1]{#1}%
\providecommand \enquote  [1]{``#1''}%
\providecommand \bibnamefont  [1]{#1}%
\providecommand \bibfnamefont [1]{#1}%
\providecommand \citenamefont [1]{#1}%
\providecommand \href@noop [0]{\@secondoftwo}%
\providecommand \href [0]{\begingroup \@sanitize@url \@href}%
\providecommand \@href[1]{\@@startlink{#1}\@@href}%
\providecommand \@@href[1]{\endgroup#1\@@endlink}%
\providecommand \@sanitize@url [0]{\catcode `\\12\catcode `\$12\catcode
  `\&12\catcode `\#12\catcode `\^12\catcode `\_12\catcode `\%12\relax}%
\providecommand \@@startlink[1]{}%
\providecommand \@@endlink[0]{}%
\providecommand \url  [0]{\begingroup\@sanitize@url \@url }%
\providecommand \@url [1]{\endgroup\@href {#1}{\urlprefix }}%
\providecommand \urlprefix  [0]{URL }%
\providecommand \Eprint [0]{\href }%
\providecommand \doibase [0]{http://dx.doi.org/}%
\providecommand \selectlanguage [0]{\@gobble}%
\providecommand \bibinfo  [0]{\@secondoftwo}%
\providecommand \bibfield  [0]{\@secondoftwo}%
\providecommand \translation [1]{[#1]}%
\providecommand \BibitemOpen [0]{}%
\providecommand \bibitemStop [0]{}%
\providecommand \bibitemNoStop [0]{.\EOS\space}%
\providecommand \EOS [0]{\spacefactor3000\relax}%
\providecommand \BibitemShut  [1]{\csname bibitem#1\endcsname}%
\let\auto@bib@innerbib\@empty
\bibitem [{\citenamefont {Bloom}\ \emph {et~al.}(2014)\citenamefont {Bloom},
  \citenamefont {Nicholson}, \citenamefont {Williams}, \citenamefont
  {Campbell}, \citenamefont {Bishof}, \citenamefont {Zhang}, \citenamefont
  {Zhang}, \citenamefont {Bromley},\ and\ \citenamefont
  {Ye}}]{bloom2014optical}%
  \BibitemOpen
  \bibfield  {author} {\bibinfo {author} {\bibfnamefont {B.}~\bibnamefont
  {Bloom}}, \bibinfo {author} {\bibfnamefont {T.}~\bibnamefont {Nicholson}},
  \bibinfo {author} {\bibfnamefont {J.}~\bibnamefont {Williams}}, \bibinfo
  {author} {\bibfnamefont {S.}~\bibnamefont {Campbell}}, \bibinfo {author}
  {\bibfnamefont {M.}~\bibnamefont {Bishof}}, \bibinfo {author} {\bibfnamefont
  {X.}~\bibnamefont {Zhang}}, \bibinfo {author} {\bibfnamefont
  {W.}~\bibnamefont {Zhang}}, \bibinfo {author} {\bibfnamefont
  {S.}~\bibnamefont {Bromley}}, \ and\ \bibinfo {author} {\bibfnamefont
  {J.}~\bibnamefont {Ye}},\ }\href@noop {} {\bibfield  {journal} {\bibinfo
  {journal} {Nature}\ }\textbf {\bibinfo {volume} {506}},\ \bibinfo {pages}
  {71} (\bibinfo {year} {2014})}\BibitemShut {NoStop}%
\bibitem [{\citenamefont {Bize}\ \emph {et~al.}(2004)\citenamefont {Bize},
  \citenamefont {Laurent}, \citenamefont {Abgrall}, \citenamefont {Marion},
  \citenamefont {Maksimovic}, \citenamefont {Cacciapuoti}, \citenamefont
  {Gr{\"u}nert}, \citenamefont {Vian}, \citenamefont {Pereira~dos Santos},
  \citenamefont {Rosenbusch} \emph {et~al.}}]{Bize2004}%
  \BibitemOpen
  \bibfield  {author} {\bibinfo {author} {\bibfnamefont {S.}~\bibnamefont
  {Bize}}, \bibinfo {author} {\bibfnamefont {P.}~\bibnamefont {Laurent}},
  \bibinfo {author} {\bibfnamefont {M.}~\bibnamefont {Abgrall}}, \bibinfo
  {author} {\bibfnamefont {H.}~\bibnamefont {Marion}}, \bibinfo {author}
  {\bibfnamefont {I.}~\bibnamefont {Maksimovic}}, \bibinfo {author}
  {\bibfnamefont {L.}~\bibnamefont {Cacciapuoti}}, \bibinfo {author}
  {\bibfnamefont {J.}~\bibnamefont {Gr{\"u}nert}}, \bibinfo {author}
  {\bibfnamefont {C.}~\bibnamefont {Vian}}, \bibinfo {author} {\bibfnamefont
  {F.}~\bibnamefont {Pereira~dos Santos}}, \bibinfo {author} {\bibfnamefont
  {P.}~\bibnamefont {Rosenbusch}},  \emph {et~al.},\ }\href@noop {} {\bibfield
  {journal} {\bibinfo  {journal} {Comptes Rendus Physique}\ }\textbf {\bibinfo
  {volume} {5}},\ \bibinfo {pages} {829} (\bibinfo {year} {2004})}\BibitemShut
  {NoStop}%
\bibitem [{\citenamefont {Laurent}\ \emph {et~al.}(2015)\citenamefont
  {Laurent}, \citenamefont {Massonnet}, \citenamefont {Cacciapuoti},\ and\
  \citenamefont {Salomon}}]{cras2014onACES}%
  \BibitemOpen
  \bibfield  {author} {\bibinfo {author} {\bibfnamefont {P.}~\bibnamefont
  {Laurent}}, \bibinfo {author} {\bibfnamefont {D.}~\bibnamefont {Massonnet}},
  \bibinfo {author} {\bibfnamefont {L.}~\bibnamefont {Cacciapuoti}}, \ and\
  \bibinfo {author} {\bibfnamefont {C.}~\bibnamefont {Salomon}},\ }\href
  {\doibase http://dx.doi.org/10.1016/j.crhy.2015.05.002} {\bibfield  {journal}
  {\bibinfo  {journal} {Comptes Rendus Physique}\ }\textbf {\bibinfo {volume}
  {16}},\ \bibinfo {pages} {540 } (\bibinfo {year} {2015})},\ \bibinfo {note}
  {the measurement of time / La mesure du temps}\BibitemShut {NoStop}%
\bibitem [{\citenamefont {Guerlin}\ \emph {et~al.}(2015)\citenamefont
  {Guerlin}, \citenamefont {Delva},\ and\ \citenamefont
  {Wolf}}]{cras2014onFundamentalTests}%
  \BibitemOpen
  \bibfield  {author} {\bibinfo {author} {\bibfnamefont {C.}~\bibnamefont
  {Guerlin}}, \bibinfo {author} {\bibfnamefont {P.}~\bibnamefont {Delva}}, \
  and\ \bibinfo {author} {\bibfnamefont {P.}~\bibnamefont {Wolf}},\ }\href
  {\doibase http://dx.doi.org/10.1016/j.crhy.2015.04.002} {\bibfield  {journal}
  {\bibinfo  {journal} {Comptes Rendus Physique}\ }\textbf {\bibinfo {volume}
  {16}},\ \bibinfo {pages} {565 } (\bibinfo {year} {2015})},\ \bibinfo {note}
  {the measurement of time / La mesure du temps}\BibitemShut {NoStop}%
\bibitem [{\citenamefont {Bongs}\ \emph {et~al.}(2015)\citenamefont {Bongs},
  \citenamefont {Singh}, \citenamefont {Smith}, \citenamefont {He},
  \citenamefont {Kock}, \citenamefont {{\'S}wierad}, \citenamefont {Hughes},
  \citenamefont {Schiller}, \citenamefont {Alighanbari}, \citenamefont
  {Origlia}, \citenamefont {Vogt}, \citenamefont {Sterr}, \citenamefont
  {Lisdat}, \citenamefont {Le~Targat}, \citenamefont {Lodewyck}, \citenamefont
  {Holleville}, \citenamefont {Venon}, \citenamefont {Bize}, \citenamefont
  {Barwood}, \citenamefont {Gill}, \citenamefont {Hill}, \citenamefont
  {Ovchinnikov}, \citenamefont {Poli}, \citenamefont {Tino}, \citenamefont
  {Stuhler},\ and\ \citenamefont {Kaenders}}]{cras2014onSpaceOpticalClocks}%
  \BibitemOpen
  \bibfield  {author} {\bibinfo {author} {\bibfnamefont {K.}~\bibnamefont
  {Bongs}}, \bibinfo {author} {\bibfnamefont {Y.}~\bibnamefont {Singh}},
  \bibinfo {author} {\bibfnamefont {L.}~\bibnamefont {Smith}}, \bibinfo
  {author} {\bibfnamefont {W.}~\bibnamefont {He}}, \bibinfo {author}
  {\bibfnamefont {O.}~\bibnamefont {Kock}}, \bibinfo {author} {\bibfnamefont
  {D.}~\bibnamefont {{\'S}wierad}}, \bibinfo {author} {\bibfnamefont
  {J.}~\bibnamefont {Hughes}}, \bibinfo {author} {\bibfnamefont
  {S.}~\bibnamefont {Schiller}}, \bibinfo {author} {\bibfnamefont
  {S.}~\bibnamefont {Alighanbari}}, \bibinfo {author} {\bibfnamefont
  {S.}~\bibnamefont {Origlia}}, \bibinfo {author} {\bibfnamefont
  {S.}~\bibnamefont {Vogt}}, \bibinfo {author} {\bibfnamefont {U.}~\bibnamefont
  {Sterr}}, \bibinfo {author} {\bibfnamefont {C.}~\bibnamefont {Lisdat}},
  \bibinfo {author} {\bibfnamefont {R.}~\bibnamefont {Le~Targat}}, \bibinfo
  {author} {\bibfnamefont {J.}~\bibnamefont {Lodewyck}}, \bibinfo {author}
  {\bibfnamefont {D.}~\bibnamefont {Holleville}}, \bibinfo {author}
  {\bibfnamefont {B.}~\bibnamefont {Venon}}, \bibinfo {author} {\bibfnamefont
  {S.}~\bibnamefont {Bize}}, \bibinfo {author} {\bibfnamefont {G.~P.}\
  \bibnamefont {Barwood}}, \bibinfo {author} {\bibfnamefont {P.}~\bibnamefont
  {Gill}}, \bibinfo {author} {\bibfnamefont {I.~R.}\ \bibnamefont {Hill}},
  \bibinfo {author} {\bibfnamefont {Y.~B.}\ \bibnamefont {Ovchinnikov}},
  \bibinfo {author} {\bibfnamefont {N.}~\bibnamefont {Poli}}, \bibinfo {author}
  {\bibfnamefont {G.~M.}\ \bibnamefont {Tino}}, \bibinfo {author}
  {\bibfnamefont {J.}~\bibnamefont {Stuhler}}, \ and\ \bibinfo {author}
  {\bibfnamefont {W.}~\bibnamefont {Kaenders}},\ }\href {\doibase
  http://dx.doi.org/10.1016/j.crhy.2015.03.009} {\bibfield  {journal} {\bibinfo
   {journal} {Comptes Rendus Physique}\ }\textbf {\bibinfo {volume} {16}},\
  \bibinfo {pages} {553 } (\bibinfo {year} {2015})},\ \bibinfo {note} {the
  measurement of time / La mesure du temps}\BibitemShut {NoStop}%
\bibitem [{\citenamefont {Lopez}\ \emph {et~al.}(2015)\citenamefont {Lopez},
  \citenamefont {K\'{e}f\'{e}lian}, \citenamefont {Jiang}, \citenamefont
  {Haboucha}, \citenamefont {Bercy}, \citenamefont {Stefani}, \citenamefont
  {Chanteau}, \citenamefont {Kanj}, \citenamefont {Rovera}, \citenamefont
  {Achkar}, \citenamefont {Chardonnet}, \citenamefont {Pottie}, \citenamefont
  {Amy-Klein},\ and\ \citenamefont {Santarelli}}]{cras2014onFiberLinks1}%
  \BibitemOpen
  \bibfield  {author} {\bibinfo {author} {\bibfnamefont {O.}~\bibnamefont
  {Lopez}}, \bibinfo {author} {\bibfnamefont {F.}~\bibnamefont
  {K\'{e}f\'{e}lian}}, \bibinfo {author} {\bibfnamefont {H.}~\bibnamefont
  {Jiang}}, \bibinfo {author} {\bibfnamefont {A.}~\bibnamefont {Haboucha}},
  \bibinfo {author} {\bibfnamefont {A.}~\bibnamefont {Bercy}}, \bibinfo
  {author} {\bibfnamefont {F.}~\bibnamefont {Stefani}}, \bibinfo {author}
  {\bibfnamefont {B.}~\bibnamefont {Chanteau}}, \bibinfo {author}
  {\bibfnamefont {A.}~\bibnamefont {Kanj}}, \bibinfo {author} {\bibfnamefont
  {D.}~\bibnamefont {Rovera}}, \bibinfo {author} {\bibfnamefont
  {J.}~\bibnamefont {Achkar}}, \bibinfo {author} {\bibfnamefont
  {C.}~\bibnamefont {Chardonnet}}, \bibinfo {author} {\bibfnamefont {P.-E.}\
  \bibnamefont {Pottie}}, \bibinfo {author} {\bibfnamefont {A.}~\bibnamefont
  {Amy-Klein}}, \ and\ \bibinfo {author} {\bibfnamefont {G.}~\bibnamefont
  {Santarelli}},\ }\href {\doibase
  http://dx.doi.org/10.1016/j.crhy.2015.04.005} {\bibfield  {journal} {\bibinfo
   {journal} {Comptes Rendus Physique}\ }\textbf {\bibinfo {volume} {16}},\
  \bibinfo {pages} {531 } (\bibinfo {year} {2015})},\ \bibinfo {note} {the
  measurement of time / La mesure du temps}\BibitemShut {NoStop}%
\bibitem [{\citenamefont {Droste}\ \emph {et~al.}(2015)\citenamefont {Droste},
  \citenamefont {Udem}, \citenamefont {Holzwarth},\ and\ \citenamefont
  {H{\"a}nsch}}]{cras2014onFiberLinks2}%
  \BibitemOpen
  \bibfield  {author} {\bibinfo {author} {\bibfnamefont {S.}~\bibnamefont
  {Droste}}, \bibinfo {author} {\bibfnamefont {T.}~\bibnamefont {Udem}},
  \bibinfo {author} {\bibfnamefont {R.}~\bibnamefont {Holzwarth}}, \ and\
  \bibinfo {author} {\bibfnamefont {T.~W.}\ \bibnamefont {H{\"a}nsch}},\ }\href
  {\doibase http://dx.doi.org/10.1016/j.crhy.2015.03.011} {\bibfield  {journal}
  {\bibinfo  {journal} {Comptes Rendus Physique}\ }\textbf {\bibinfo {volume}
  {16}},\ \bibinfo {pages} {524 } (\bibinfo {year} {2015})},\ \bibinfo {note}
  {the measurement of time / La mesure du temps}\BibitemShut {NoStop}%
\bibitem [{\citenamefont {Gu{\'e}na}\ \emph {et~al.}(2012)\citenamefont
  {Gu{\'e}na}, \citenamefont {Abgrall}, \citenamefont {Rovera}, \citenamefont
  {Laurent}, \citenamefont {Chupin}, \citenamefont {Lours}, \citenamefont
  {Santarelli}, \citenamefont {Rosenbusch}, \citenamefont {Tobar},
  \citenamefont {Li} \emph {et~al.}}]{guena2012}%
  \BibitemOpen
  \bibfield  {author} {\bibinfo {author} {\bibfnamefont {J.}~\bibnamefont
  {Gu{\'e}na}}, \bibinfo {author} {\bibfnamefont {M.}~\bibnamefont {Abgrall}},
  \bibinfo {author} {\bibfnamefont {D.}~\bibnamefont {Rovera}}, \bibinfo
  {author} {\bibfnamefont {P.}~\bibnamefont {Laurent}}, \bibinfo {author}
  {\bibfnamefont {B.}~\bibnamefont {Chupin}}, \bibinfo {author} {\bibfnamefont
  {M.}~\bibnamefont {Lours}}, \bibinfo {author} {\bibfnamefont
  {G.}~\bibnamefont {Santarelli}}, \bibinfo {author} {\bibfnamefont
  {P.}~\bibnamefont {Rosenbusch}}, \bibinfo {author} {\bibfnamefont {M.~E.}\
  \bibnamefont {Tobar}}, \bibinfo {author} {\bibfnamefont {R.}~\bibnamefont
  {Li}},  \emph {et~al.},\ }\href@noop {} {\bibfield  {journal} {\bibinfo
  {journal} {IEEE transactions on Ultrasonics, Ferroelectrics and Frequency
  Control}\ }\textbf {\bibinfo {volume} {59}},\ \bibinfo {pages} {391}
  (\bibinfo {year} {2012})}\BibitemShut {NoStop}%
\bibitem [{\citenamefont {Li}\ and\ \citenamefont {Gibble}(2004)}]{li2004}%
  \BibitemOpen
  \bibfield  {author} {\bibinfo {author} {\bibfnamefont {R.}~\bibnamefont
  {Li}}\ and\ \bibinfo {author} {\bibfnamefont {K.}~\bibnamefont {Gibble}},\
  }\href@noop {} {\bibfield  {journal} {\bibinfo  {journal} {Metrologia}\
  }\textbf {\bibinfo {volume} {41}},\ \bibinfo {pages} {376} (\bibinfo {year}
  {2004})}\BibitemShut {NoStop}%
\bibitem [{\citenamefont {Li}\ and\ \citenamefont {Gibble}(2010)}]{li2010}%
  \BibitemOpen
  \bibfield  {author} {\bibinfo {author} {\bibfnamefont {R.}~\bibnamefont
  {Li}}\ and\ \bibinfo {author} {\bibfnamefont {K.}~\bibnamefont {Gibble}},\
  }\href@noop {} {\bibfield  {journal} {\bibinfo  {journal} {Metrologia}\
  }\textbf {\bibinfo {volume} {47}},\ \bibinfo {pages} {534} (\bibinfo {year}
  {2010})}\BibitemShut {NoStop}%
\bibitem [{\citenamefont {Gu{\'e}na}\ \emph {et~al.}(2011)\citenamefont
  {Gu{\'e}na}, \citenamefont {Li}, \citenamefont {Gibble}, \citenamefont
  {Bize},\ and\ \citenamefont {Clairon}}]{guena2011}%
  \BibitemOpen
  \bibfield  {author} {\bibinfo {author} {\bibfnamefont {J.}~\bibnamefont
  {Gu{\'e}na}}, \bibinfo {author} {\bibfnamefont {R.}~\bibnamefont {Li}},
  \bibinfo {author} {\bibfnamefont {K.}~\bibnamefont {Gibble}}, \bibinfo
  {author} {\bibfnamefont {S.}~\bibnamefont {Bize}}, \ and\ \bibinfo {author}
  {\bibfnamefont {A.}~\bibnamefont {Clairon}},\ }\href@noop {} {\bibfield
  {journal} {\bibinfo  {journal} {Phys. Rev. Lett.}\ }\textbf {\bibinfo
  {volume} {106}},\ \bibinfo {pages} {130801} (\bibinfo {year}
  {2011})}\BibitemShut {NoStop}%
\bibitem [{\citenamefont {Li}\ \emph {et~al.}(2011)\citenamefont {Li},
  \citenamefont {Gibble},\ and\ \citenamefont {Szymaniec}}]{li2011}%
  \BibitemOpen
  \bibfield  {author} {\bibinfo {author} {\bibfnamefont {R.}~\bibnamefont
  {Li}}, \bibinfo {author} {\bibfnamefont {K.}~\bibnamefont {Gibble}}, \ and\
  \bibinfo {author} {\bibfnamefont {K.}~\bibnamefont {Szymaniec}},\ }\href@noop
  {} {\bibfield  {journal} {\bibinfo  {journal} {Metrologia}\ }\textbf
  {\bibinfo {volume} {48}},\ \bibinfo {pages} {283} (\bibinfo {year}
  {2011})}\BibitemShut {NoStop}%
\bibitem [{\citenamefont {Weyers}\ \emph {et~al.}(2012)\citenamefont {Weyers},
  \citenamefont {Gerginov}, \citenamefont {Nemitz}, \citenamefont {Li},\ and\
  \citenamefont {Gibble}}]{weyers2012}%
  \BibitemOpen
  \bibfield  {author} {\bibinfo {author} {\bibfnamefont {S.}~\bibnamefont
  {Weyers}}, \bibinfo {author} {\bibfnamefont {V.}~\bibnamefont {Gerginov}},
  \bibinfo {author} {\bibfnamefont {N.}~\bibnamefont {Nemitz}}, \bibinfo
  {author} {\bibfnamefont {R.}~\bibnamefont {Li}}, \ and\ \bibinfo {author}
  {\bibfnamefont {K.}~\bibnamefont {Gibble}},\ }\href@noop {} {\bibfield
  {journal} {\bibinfo  {journal} {Metrologia}\ }\textbf {\bibinfo {volume}
  {49}},\ \bibinfo {pages} {82} (\bibinfo {year} {2012})}\BibitemShut {NoStop}%
\bibitem [{\citenamefont {Bord\'e}(2002)}]{Borde2002}%
  \BibitemOpen
  \bibfield  {author} {\bibinfo {author} {\bibfnamefont {C.~J.}\ \bibnamefont
  {Bord\'e}},\ }\href@noop {} {\bibfield  {journal} {\bibinfo  {journal}
  {Metrologia}\ }\textbf {\bibinfo {volume} {39}},\ \bibinfo {pages} {435}
  (\bibinfo {year} {2002})}\BibitemShut {NoStop}%
\bibitem [{\citenamefont {Wolf}\ and\ \citenamefont
  {Bord{\'e}}(2004)}]{Wolf2004}%
  \BibitemOpen
  \bibfield  {author} {\bibinfo {author} {\bibfnamefont {P.}~\bibnamefont
  {Wolf}}\ and\ \bibinfo {author} {\bibfnamefont {C.~J.}\ \bibnamefont
  {Bord{\'e}}},\ }\href@noop {} {\bibfield  {journal} {\bibinfo  {journal}
  {arxiv:quant-ph/0403194v1}\ } (\bibinfo {year} {2004})}\BibitemShut {NoStop}%
\bibitem [{\citenamefont {Gibble}(2006)}]{Gibble2006}%
  \BibitemOpen
  \bibfield  {author} {\bibinfo {author} {\bibfnamefont {K.}~\bibnamefont
  {Gibble}},\ }\href@noop {} {\bibfield  {journal} {\bibinfo  {journal} {Phys.
  Rev. Lett.}\ }\textbf {\bibinfo {volume} {97}},\ \bibinfo {pages} {073002}
  (\bibinfo {year} {2006})}\BibitemShut {NoStop}%
\bibitem [{\citenamefont {Simon}\ \emph {et~al.}(1998)\citenamefont {Simon},
  \citenamefont {Laurent},\ and\ \citenamefont {Clairon}}]{simon1998}%
  \BibitemOpen
  \bibfield  {author} {\bibinfo {author} {\bibfnamefont {E.}~\bibnamefont
  {Simon}}, \bibinfo {author} {\bibfnamefont {P.}~\bibnamefont {Laurent}}, \
  and\ \bibinfo {author} {\bibfnamefont {A.}~\bibnamefont {Clairon}},\
  }\href@noop {} {\bibfield  {journal} {\bibinfo  {journal} {Phys. Rev. A}\
  }\textbf {\bibinfo {volume} {57}},\ \bibinfo {pages} {436} (\bibinfo {year}
  {1998})}\BibitemShut {NoStop}%
\bibitem [{\citenamefont {Rosenbusch}\ \emph {et~al.}(2007)\citenamefont
  {Rosenbusch}, \citenamefont {Zhang},\ and\ \citenamefont
  {Clairon}}]{rosenbusch2007}%
  \BibitemOpen
  \bibfield  {author} {\bibinfo {author} {\bibfnamefont {P.}~\bibnamefont
  {Rosenbusch}}, \bibinfo {author} {\bibfnamefont {S.}~\bibnamefont {Zhang}}, \
  and\ \bibinfo {author} {\bibfnamefont {C.}~\bibnamefont {Clairon}},\ }in\
  \href@noop {} {\emph {\bibinfo {booktitle} {proceedings of the 2007 IEEE
  International Frequency Control Symposium joint with the 21$^{\mathrm{st}}$
  European Frequency and Time Forum}}}\ (\bibinfo {year} {2007})\ pp.\ \bibinfo
  {pages} {1060--1063}\BibitemShut {NoStop}%
\bibitem [{\citenamefont {Santarelli}\ \emph {et~al.}(2009)\citenamefont
  {Santarelli}, \citenamefont {Governatori}, \citenamefont {Chambon},
  \citenamefont {Lours}, \citenamefont {Rosenbusch}, \citenamefont {Gu\'ena},
  \citenamefont {Chapelet}, \citenamefont {Bize}, \citenamefont {Tobar},
  \citenamefont {Laurent} \emph {et~al.}}]{santarelli2009}%
  \BibitemOpen
  \bibfield  {author} {\bibinfo {author} {\bibfnamefont {G.}~\bibnamefont
  {Santarelli}}, \bibinfo {author} {\bibfnamefont {G.}~\bibnamefont
  {Governatori}}, \bibinfo {author} {\bibfnamefont {D.}~\bibnamefont
  {Chambon}}, \bibinfo {author} {\bibfnamefont {M.}~\bibnamefont {Lours}},
  \bibinfo {author} {\bibfnamefont {P.}~\bibnamefont {Rosenbusch}}, \bibinfo
  {author} {\bibfnamefont {J.}~\bibnamefont {Gu\'ena}}, \bibinfo {author}
  {\bibfnamefont {F.}~\bibnamefont {Chapelet}}, \bibinfo {author}
  {\bibfnamefont {S.}~\bibnamefont {Bize}}, \bibinfo {author} {\bibfnamefont
  {M.~E.}\ \bibnamefont {Tobar}}, \bibinfo {author} {\bibfnamefont
  {P.}~\bibnamefont {Laurent}},  \emph {et~al.},\ }\href@noop {} {\bibfield
  {journal} {\bibinfo  {journal} {IEEE transactions on Ultrasonics,
  Ferroelectrics and Frequency Control}\ }\textbf {\bibinfo {volume} {56}},\
  \bibinfo {pages} {1319} (\bibinfo {year} {2009})}\BibitemShut {NoStop}%
\bibitem [{\citenamefont {Gu\'ena}\ \emph {et~al.}(2010)\citenamefont
  {Gu\'ena}, \citenamefont {Rosenbusch}, \citenamefont {Laurent}, \citenamefont
  {Abgrall}, \citenamefont {Rovera}, \citenamefont {Santarelli}, \citenamefont
  {Tobar}, \citenamefont {Bize},\ and\ \citenamefont {Clairon}}]{guena2010}%
  \BibitemOpen
  \bibfield  {author} {\bibinfo {author} {\bibfnamefont {J.}~\bibnamefont
  {Gu\'ena}}, \bibinfo {author} {\bibfnamefont {P.}~\bibnamefont {Rosenbusch}},
  \bibinfo {author} {\bibfnamefont {P.}~\bibnamefont {Laurent}}, \bibinfo
  {author} {\bibfnamefont {M.}~\bibnamefont {Abgrall}}, \bibinfo {author}
  {\bibfnamefont {D.}~\bibnamefont {Rovera}}, \bibinfo {author} {\bibfnamefont
  {G.}~\bibnamefont {Santarelli}}, \bibinfo {author} {\bibfnamefont {M.~E.}\
  \bibnamefont {Tobar}}, \bibinfo {author} {\bibfnamefont {S.}~\bibnamefont
  {Bize}}, \ and\ \bibinfo {author} {\bibfnamefont {C.}~\bibnamefont
  {Clairon}},\ }\href@noop {} {\bibfield  {journal} {\bibinfo  {journal} {IEEE
  transactions on Ultrasonics, Ferroelectrics and Frequency Control}\ }\textbf
  {\bibinfo {volume} {57}},\ \bibinfo {pages} {647} (\bibinfo {year}
  {2010})}\BibitemShut {NoStop}%
\bibitem [{\citenamefont {Katori}(2001)}]{Katori2001}%
  \BibitemOpen
  \bibfield  {author} {\bibinfo {author} {\bibfnamefont {H.}~\bibnamefont
  {Katori}},\ }in\ \href@noop {} {\emph {\bibinfo {booktitle} {proceedings of
  the 6$^{\mathrm{th}}$ Symposium on Frequency Standards and Metrology}}}\
  (\bibinfo  {publisher} {World scientific, Singapore},\ \bibinfo {year}
  {2001})\ p.\ \bibinfo {pages} {323}\BibitemShut {NoStop}%
\bibitem [{\citenamefont {Katori}\ \emph {et~al.}(2003)\citenamefont {Katori},
  \citenamefont {Takamoto}, \citenamefont {Pal\char39{}chikov},\ and\
  \citenamefont {Ovsiannikov}}]{Katori2003a}%
  \BibitemOpen
  \bibfield  {author} {\bibinfo {author} {\bibfnamefont {H.}~\bibnamefont
  {Katori}}, \bibinfo {author} {\bibfnamefont {M.}~\bibnamefont {Takamoto}},
  \bibinfo {author} {\bibfnamefont {V.~G.}\ \bibnamefont {Pal\char39{}chikov}},
  \ and\ \bibinfo {author} {\bibfnamefont {V.~D.}\ \bibnamefont
  {Ovsiannikov}},\ }\href@noop {} {\bibfield  {journal} {\bibinfo  {journal}
  {Phys. Rev. Lett.}\ }\textbf {\bibinfo {volume} {91}},\ \bibinfo {pages}
  {173005} (\bibinfo {year} {2003})}\BibitemShut {NoStop}%
\bibitem [{\citenamefont {Le~Targat}\ \emph {et~al.}(2013)\citenamefont
  {Le~Targat}, \citenamefont {Lorini}, \citenamefont {Le~Coq}, \citenamefont
  {Zawada}, \citenamefont {Gu{\'e}na}, \citenamefont {Abgrall}, \citenamefont
  {Gurov}, \citenamefont {Rosenbusch}, \citenamefont {Rovera}, \citenamefont
  {Nag{\'o}rny} \emph {et~al.}}]{letargat2013}%
  \BibitemOpen
  \bibfield  {author} {\bibinfo {author} {\bibfnamefont {R.}~\bibnamefont
  {Le~Targat}}, \bibinfo {author} {\bibfnamefont {L.}~\bibnamefont {Lorini}},
  \bibinfo {author} {\bibfnamefont {Y.}~\bibnamefont {Le~Coq}}, \bibinfo
  {author} {\bibfnamefont {M.}~\bibnamefont {Zawada}}, \bibinfo {author}
  {\bibfnamefont {J.}~\bibnamefont {Gu{\'e}na}}, \bibinfo {author}
  {\bibfnamefont {M.}~\bibnamefont {Abgrall}}, \bibinfo {author} {\bibfnamefont
  {M.}~\bibnamefont {Gurov}}, \bibinfo {author} {\bibfnamefont
  {P.}~\bibnamefont {Rosenbusch}}, \bibinfo {author} {\bibfnamefont
  {D.}~\bibnamefont {Rovera}}, \bibinfo {author} {\bibfnamefont
  {B.}~\bibnamefont {Nag{\'o}rny}},  \emph {et~al.},\ }\href@noop {} {\bibfield
   {journal} {\bibinfo  {journal} {Nature Communications}\ }\textbf {\bibinfo
  {volume} {4}} (\bibinfo {year} {2013})}\BibitemShut {NoStop}%
\bibitem [{\citenamefont {Hinkley}\ \emph {et~al.}(2013)\citenamefont
  {Hinkley}, \citenamefont {Sherman}, \citenamefont {Phillips}, \citenamefont
  {Schioppo}, \citenamefont {Lemke}, \citenamefont {Beloy}, \citenamefont
  {Pizzocaro}, \citenamefont {Oates},\ and\ \citenamefont
  {Ludlow}}]{hinkley_atomic_2013}%
  \BibitemOpen
  \bibfield  {author} {\bibinfo {author} {\bibfnamefont {N.}~\bibnamefont
  {Hinkley}}, \bibinfo {author} {\bibfnamefont {J.~A.}\ \bibnamefont
  {Sherman}}, \bibinfo {author} {\bibfnamefont {N.~B.}\ \bibnamefont
  {Phillips}}, \bibinfo {author} {\bibfnamefont {M.}~\bibnamefont {Schioppo}},
  \bibinfo {author} {\bibfnamefont {N.~D.}\ \bibnamefont {Lemke}}, \bibinfo
  {author} {\bibfnamefont {K.}~\bibnamefont {Beloy}}, \bibinfo {author}
  {\bibfnamefont {M.}~\bibnamefont {Pizzocaro}}, \bibinfo {author}
  {\bibfnamefont {C.~W.}\ \bibnamefont {Oates}}, \ and\ \bibinfo {author}
  {\bibfnamefont {A.~D.}\ \bibnamefont {Ludlow}},\ }\href@noop {} {\bibfield
  {journal} {\bibinfo  {journal} {Science}\ }\textbf {\bibinfo {volume}
  {341}},\ \bibinfo {pages} {1215} (\bibinfo {year} {2013})}\BibitemShut
  {NoStop}%
\bibitem [{\citenamefont {Falke}\ \emph {et~al.}(2014)\citenamefont {Falke},
  \citenamefont {Lemke}, \citenamefont {Grebing}, \citenamefont {Lipphardt},
  \citenamefont {Weyers}, \citenamefont {Gerginov}, \citenamefont {Huntemann},
  \citenamefont {Hagemann}, \citenamefont {Al-Masoudi}, \citenamefont
  {H\"afner}, \citenamefont {Vogt}, \citenamefont {Sterr},\ and\ \citenamefont
  {Lisdat}}]{falke2014}%
  \BibitemOpen
  \bibfield  {author} {\bibinfo {author} {\bibfnamefont {S.}~\bibnamefont
  {Falke}}, \bibinfo {author} {\bibfnamefont {N.}~\bibnamefont {Lemke}},
  \bibinfo {author} {\bibfnamefont {C.}~\bibnamefont {Grebing}}, \bibinfo
  {author} {\bibfnamefont {B.}~\bibnamefont {Lipphardt}}, \bibinfo {author}
  {\bibfnamefont {S.}~\bibnamefont {Weyers}}, \bibinfo {author} {\bibfnamefont
  {V.}~\bibnamefont {Gerginov}}, \bibinfo {author} {\bibfnamefont
  {N.}~\bibnamefont {Huntemann}}, \bibinfo {author} {\bibfnamefont
  {C.}~\bibnamefont {Hagemann}}, \bibinfo {author} {\bibfnamefont
  {A.}~\bibnamefont {Al-Masoudi}}, \bibinfo {author} {\bibfnamefont
  {S.}~\bibnamefont {H\"afner}}, \bibinfo {author} {\bibfnamefont
  {S.}~\bibnamefont {Vogt}}, \bibinfo {author} {\bibfnamefont {U.}~\bibnamefont
  {Sterr}}, \ and\ \bibinfo {author} {\bibfnamefont {C.}~\bibnamefont
  {Lisdat}},\ }\href@noop {} {\bibfield  {journal} {\bibinfo  {journal} {New
  Journal of Physics}\ }\textbf {\bibinfo {volume} {16}},\ \bibinfo {pages}
  {073023} (\bibinfo {year} {2014})}\BibitemShut {NoStop}%
\bibitem [{\citenamefont {Ushijima}\ \emph {et~al.}(2015)\citenamefont
  {Ushijima}, \citenamefont {Takamoto}, \citenamefont {Das}, \citenamefont
  {Ohkubo},\ and\ \citenamefont {Katori}}]{ushijima2014cryogenic}%
  \BibitemOpen
  \bibfield  {author} {\bibinfo {author} {\bibfnamefont {I.}~\bibnamefont
  {Ushijima}}, \bibinfo {author} {\bibfnamefont {M.}~\bibnamefont {Takamoto}},
  \bibinfo {author} {\bibfnamefont {M.}~\bibnamefont {Das}}, \bibinfo {author}
  {\bibfnamefont {T.}~\bibnamefont {Ohkubo}}, \ and\ \bibinfo {author}
  {\bibfnamefont {H.}~\bibnamefont {Katori}},\ }\href@noop {} {\bibfield
  {journal} {\bibinfo  {journal} {Nature Photonics}\ }\textbf {\bibinfo
  {volume} {9}},\ \bibinfo {pages} {185} (\bibinfo {year} {2015})}\BibitemShut
  {NoStop}%
\bibitem [{\citenamefont {Brusch}\ \emph {et~al.}(2006)\citenamefont {Brusch},
  \citenamefont {Le~Targat}, \citenamefont {Baillard}, \citenamefont
  {Fouch\'e},\ and\ \citenamefont {Lemonde}}]{Brusch2006a}%
  \BibitemOpen
  \bibfield  {author} {\bibinfo {author} {\bibfnamefont {A.}~\bibnamefont
  {Brusch}}, \bibinfo {author} {\bibfnamefont {R.}~\bibnamefont {Le~Targat}},
  \bibinfo {author} {\bibfnamefont {X.}~\bibnamefont {Baillard}}, \bibinfo
  {author} {\bibfnamefont {M.}~\bibnamefont {Fouch\'e}}, \ and\ \bibinfo
  {author} {\bibfnamefont {P.}~\bibnamefont {Lemonde}},\ }\href@noop {}
  {\bibfield  {journal} {\bibinfo  {journal} {Phys. Rev. Lett.}\ }\textbf
  {\bibinfo {volume} {96}},\ \bibinfo {pages} {103003} (\bibinfo {year}
  {2006})}\BibitemShut {NoStop}%
\bibitem [{\citenamefont {Westergaard}\ \emph {et~al.}(2011)\citenamefont
  {Westergaard}, \citenamefont {Lodewyck}, \citenamefont {Lorini},
  \citenamefont {Lecallier}, \citenamefont {Burt}, \citenamefont {Zawada},
  \citenamefont {Millo},\ and\ \citenamefont
  {Lemonde}}]{westergaard2011lattice}%
  \BibitemOpen
  \bibfield  {author} {\bibinfo {author} {\bibfnamefont {P.~G.}\ \bibnamefont
  {Westergaard}}, \bibinfo {author} {\bibfnamefont {J.}~\bibnamefont
  {Lodewyck}}, \bibinfo {author} {\bibfnamefont {L.}~\bibnamefont {Lorini}},
  \bibinfo {author} {\bibfnamefont {A.}~\bibnamefont {Lecallier}}, \bibinfo
  {author} {\bibfnamefont {E.}~\bibnamefont {Burt}}, \bibinfo {author}
  {\bibfnamefont {M.}~\bibnamefont {Zawada}}, \bibinfo {author} {\bibfnamefont
  {J.}~\bibnamefont {Millo}}, \ and\ \bibinfo {author} {\bibfnamefont
  {P.}~\bibnamefont {Lemonde}},\ }\href@noop {} {\bibfield  {journal} {\bibinfo
   {journal} {Phys. Rev. Lett.}\ }\textbf {\bibinfo {volume} {106}},\ \bibinfo
  {pages} {210801} (\bibinfo {year} {2011})}\BibitemShut {NoStop}%
\bibitem [{\citenamefont {Petersen}\ \emph
  {et~al.}(2008{\natexlab{a}})\citenamefont {Petersen}, \citenamefont {Millo},
  \citenamefont {Magalhaes}, \citenamefont {Mandache}, \citenamefont {Dawkins},
  \citenamefont {Chicireanu}, \citenamefont {Le~Coq}, \citenamefont {Acef},
  \citenamefont {Santarelli}, \citenamefont {Clairon},\ and\ \citenamefont
  {Bize}}]{Petersen2008b}%
  \BibitemOpen
  \bibfield  {author} {\bibinfo {author} {\bibfnamefont {M.}~\bibnamefont
  {Petersen}}, \bibinfo {author} {\bibfnamefont {J.}~\bibnamefont {Millo}},
  \bibinfo {author} {\bibfnamefont {D.}~\bibnamefont {Magalhaes}}, \bibinfo
  {author} {\bibfnamefont {C.}~\bibnamefont {Mandache}}, \bibinfo {author}
  {\bibfnamefont {S.}~\bibnamefont {Dawkins}}, \bibinfo {author} {\bibfnamefont
  {R.}~\bibnamefont {Chicireanu}}, \bibinfo {author} {\bibfnamefont
  {Y.}~\bibnamefont {Le~Coq}}, \bibinfo {author} {\bibfnamefont
  {O.}~\bibnamefont {Acef}}, \bibinfo {author} {\bibfnamefont {G.}~\bibnamefont
  {Santarelli}}, \bibinfo {author} {\bibfnamefont {A.}~\bibnamefont {Clairon}},
  \ and\ \bibinfo {author} {\bibfnamefont {S.}~\bibnamefont {Bize}},\ }in\
  \href@noop {} {\emph {\bibinfo {booktitle} {proceedings of the 2008 IEEE
  International Frequency Control Symposium}}}\ (\bibinfo {year} {2008})\ pp.\
  \bibinfo {pages} {451 --454}\BibitemShut {NoStop}%
\bibitem [{\citenamefont {Hachisu}\ \emph {et~al.}(2008)\citenamefont
  {Hachisu}, \citenamefont {Miyagishi}, \citenamefont {Porsev}, \citenamefont
  {Derevianko}, \citenamefont {Ovsiannikov}, \citenamefont {Pal'chikov},
  \citenamefont {Takamoto},\ and\ \citenamefont {Katori}}]{Hachisu2008}%
  \BibitemOpen
  \bibfield  {author} {\bibinfo {author} {\bibfnamefont {H.}~\bibnamefont
  {Hachisu}}, \bibinfo {author} {\bibfnamefont {K.}~\bibnamefont {Miyagishi}},
  \bibinfo {author} {\bibfnamefont {S.~G.}\ \bibnamefont {Porsev}}, \bibinfo
  {author} {\bibfnamefont {A.}~\bibnamefont {Derevianko}}, \bibinfo {author}
  {\bibfnamefont {V.~D.}\ \bibnamefont {Ovsiannikov}}, \bibinfo {author}
  {\bibfnamefont {V.~G.}\ \bibnamefont {Pal'chikov}}, \bibinfo {author}
  {\bibfnamefont {M.}~\bibnamefont {Takamoto}}, \ and\ \bibinfo {author}
  {\bibfnamefont {H.}~\bibnamefont {Katori}},\ }\href@noop {} {\bibfield
  {journal} {\bibinfo  {journal} {Phys. Rev. Lett.}\ }\textbf {\bibinfo
  {volume} {100}},\ \bibinfo {pages} {053001} (\bibinfo {year}
  {2008})}\BibitemShut {NoStop}%
\bibitem [{\citenamefont {McFerran}\ \emph {et~al.}(2010)\citenamefont
  {McFerran}, \citenamefont {Yi}, \citenamefont {Mejri},\ and\ \citenamefont
  {Bize}}]{McFerran2010}%
  \BibitemOpen
  \bibfield  {author} {\bibinfo {author} {\bibfnamefont {J.~J.}\ \bibnamefont
  {McFerran}}, \bibinfo {author} {\bibfnamefont {L.}~\bibnamefont {Yi}},
  \bibinfo {author} {\bibfnamefont {S.}~\bibnamefont {Mejri}}, \ and\ \bibinfo
  {author} {\bibfnamefont {S.}~\bibnamefont {Bize}},\ }\href@noop {} {\bibfield
   {journal} {\bibinfo  {journal} {Opt. Lett.}\ }\textbf {\bibinfo {volume}
  {35}},\ \bibinfo {pages} {3078} (\bibinfo {year} {2010})}\BibitemShut
  {NoStop}%
\bibitem [{\citenamefont {Millo}\ \emph
  {et~al.}(2009{\natexlab{a}})\citenamefont {Millo}, \citenamefont {Magalhaes},
  \citenamefont {Mandache}, \citenamefont {Le~Coq}, \citenamefont {English},
  \citenamefont {Westergaard}, \citenamefont {Lodewyck}, \citenamefont {Bize},
  \citenamefont {Lemonde},\ and\ \citenamefont {Santarelli}}]{Millo2009b}%
  \BibitemOpen
  \bibfield  {author} {\bibinfo {author} {\bibfnamefont {J.}~\bibnamefont
  {Millo}}, \bibinfo {author} {\bibfnamefont {D.~V.}\ \bibnamefont
  {Magalhaes}}, \bibinfo {author} {\bibfnamefont {C.}~\bibnamefont {Mandache}},
  \bibinfo {author} {\bibfnamefont {Y.}~\bibnamefont {Le~Coq}}, \bibinfo
  {author} {\bibfnamefont {E.~M.~L.}\ \bibnamefont {English}}, \bibinfo
  {author} {\bibfnamefont {P.~G.}\ \bibnamefont {Westergaard}}, \bibinfo
  {author} {\bibfnamefont {J.}~\bibnamefont {Lodewyck}}, \bibinfo {author}
  {\bibfnamefont {S.}~\bibnamefont {Bize}}, \bibinfo {author} {\bibfnamefont
  {P.}~\bibnamefont {Lemonde}}, \ and\ \bibinfo {author} {\bibfnamefont
  {G.}~\bibnamefont {Santarelli}},\ }\href@noop {} {\bibfield  {journal}
  {\bibinfo  {journal} {Phys. Rev. A}\ }\textbf {\bibinfo {volume} {79}},\
  \bibinfo {pages} {053829} (\bibinfo {year} {2009}{\natexlab{a}})}\BibitemShut
  {NoStop}%
\bibitem [{\citenamefont {Dawkins}\ \emph {et~al.}(2010)\citenamefont
  {Dawkins}, \citenamefont {Chicireanu}, \citenamefont {Petersen},
  \citenamefont {Millo}, \citenamefont {Magalhães}, \citenamefont {Mandache},
  \citenamefont {{Le Coq}},\ and\ \citenamefont {Bize}}]{Dawkins2010}%
  \BibitemOpen
  \bibfield  {author} {\bibinfo {author} {\bibfnamefont {S.}~\bibnamefont
  {Dawkins}}, \bibinfo {author} {\bibfnamefont {R.}~\bibnamefont {Chicireanu}},
  \bibinfo {author} {\bibfnamefont {M.}~\bibnamefont {Petersen}}, \bibinfo
  {author} {\bibfnamefont {J.}~\bibnamefont {Millo}}, \bibinfo {author}
  {\bibfnamefont {D.}~\bibnamefont {Magalhães}}, \bibinfo {author}
  {\bibfnamefont {C.}~\bibnamefont {Mandache}}, \bibinfo {author}
  {\bibfnamefont {Y.}~\bibnamefont {{Le Coq}}}, \ and\ \bibinfo {author}
  {\bibfnamefont {S.}~\bibnamefont {Bize}},\ }\href@noop {} {\bibfield
  {journal} {\bibinfo  {journal} {Appl. Phys. B: Lasers and Optics}\ }\textbf
  {\bibinfo {volume} {99}},\ \bibinfo {pages} {41} (\bibinfo {year}
  {2010})}\BibitemShut {NoStop}%
\bibitem [{\citenamefont {Petersen}\ \emph
  {et~al.}(2008{\natexlab{b}})\citenamefont {Petersen}, \citenamefont
  {Chicireanu}, \citenamefont {Dawkins}, \citenamefont {Magalhaes},
  \citenamefont {Mandache}, \citenamefont {Le~Coq}, \citenamefont {Clairon},\
  and\ \citenamefont {Bize}}]{Petersen2008a}%
  \BibitemOpen
  \bibfield  {author} {\bibinfo {author} {\bibfnamefont {M.}~\bibnamefont
  {Petersen}}, \bibinfo {author} {\bibfnamefont {R.}~\bibnamefont
  {Chicireanu}}, \bibinfo {author} {\bibfnamefont {S.~T.}\ \bibnamefont
  {Dawkins}}, \bibinfo {author} {\bibfnamefont {D.~V.}\ \bibnamefont
  {Magalhaes}}, \bibinfo {author} {\bibfnamefont {C.}~\bibnamefont {Mandache}},
  \bibinfo {author} {\bibfnamefont {Y.}~\bibnamefont {Le~Coq}}, \bibinfo
  {author} {\bibfnamefont {A.}~\bibnamefont {Clairon}}, \ and\ \bibinfo
  {author} {\bibfnamefont {S.}~\bibnamefont {Bize}},\ }\href@noop {} {\bibfield
   {journal} {\bibinfo  {journal} {Phys. Rev. Lett.}\ }\textbf {\bibinfo
  {volume} {101}},\ \bibinfo {pages} {183004} (\bibinfo {year}
  {2008}{\natexlab{b}})}\BibitemShut {NoStop}%
\bibitem [{\citenamefont {Yi}\ \emph {et~al.}(2011)\citenamefont {Yi},
  \citenamefont {Mejri}, \citenamefont {McFerran}, \citenamefont {Le~Coq},\
  and\ \citenamefont {Bize}}]{Yi2011}%
  \BibitemOpen
  \bibfield  {author} {\bibinfo {author} {\bibfnamefont {L.}~\bibnamefont
  {Yi}}, \bibinfo {author} {\bibfnamefont {S.}~\bibnamefont {Mejri}}, \bibinfo
  {author} {\bibfnamefont {J.~J.}\ \bibnamefont {McFerran}}, \bibinfo {author}
  {\bibfnamefont {Y.}~\bibnamefont {Le~Coq}}, \ and\ \bibinfo {author}
  {\bibfnamefont {S.}~\bibnamefont {Bize}},\ }\href@noop {} {\bibfield
  {journal} {\bibinfo  {journal} {Phys. Rev. Lett.}\ }\textbf {\bibinfo
  {volume} {106}},\ \bibinfo {pages} {073005} (\bibinfo {year}
  {2011})}\BibitemShut {NoStop}%
\bibitem [{\citenamefont {McFerran}\ \emph {et~al.}(2012)\citenamefont
  {McFerran}, \citenamefont {Yi}, \citenamefont {Mejri}, \citenamefont
  {Di~Manno}, \citenamefont {Zhang}, \citenamefont {Gu\'ena}, \citenamefont
  {Le~Coq},\ and\ \citenamefont {Bize}}]{McFerran2012}%
  \BibitemOpen
  \bibfield  {author} {\bibinfo {author} {\bibfnamefont {J.~J.}\ \bibnamefont
  {McFerran}}, \bibinfo {author} {\bibfnamefont {L.}~\bibnamefont {Yi}},
  \bibinfo {author} {\bibfnamefont {S.}~\bibnamefont {Mejri}}, \bibinfo
  {author} {\bibfnamefont {S.}~\bibnamefont {Di~Manno}}, \bibinfo {author}
  {\bibfnamefont {W.}~\bibnamefont {Zhang}}, \bibinfo {author} {\bibfnamefont
  {J.}~\bibnamefont {Gu\'ena}}, \bibinfo {author} {\bibfnamefont
  {Y.}~\bibnamefont {Le~Coq}}, \ and\ \bibinfo {author} {\bibfnamefont
  {S.}~\bibnamefont {Bize}},\ }\href@noop {} {\bibfield  {journal} {\bibinfo
  {journal} {Phys. Rev. Lett.}\ }\textbf {\bibinfo {volume} {108}},\ \bibinfo
  {pages} {183004} (\bibinfo {year} {2012})}\BibitemShut {NoStop}%
\bibitem [{\citenamefont {Millo}\ \emph
  {et~al.}(2009{\natexlab{b}})\citenamefont {Millo}, \citenamefont {Abgrall},
  \citenamefont {Lours}, \citenamefont {English}, \citenamefont {Jiang},
  \citenamefont {Gu{\'e}na}, \citenamefont {Clairon}, \citenamefont {Tobar},
  \citenamefont {Bize}, \citenamefont {Le~Coq},\ and\ \citenamefont
  {Santarelli}}]{MilloAPL2009}%
  \BibitemOpen
  \bibfield  {author} {\bibinfo {author} {\bibfnamefont {J.}~\bibnamefont
  {Millo}}, \bibinfo {author} {\bibfnamefont {M.}~\bibnamefont {Abgrall}},
  \bibinfo {author} {\bibfnamefont {M.}~\bibnamefont {Lours}}, \bibinfo
  {author} {\bibfnamefont {E.}~\bibnamefont {English}}, \bibinfo {author}
  {\bibfnamefont {H.}~\bibnamefont {Jiang}}, \bibinfo {author} {\bibfnamefont
  {J.}~\bibnamefont {Gu{\'e}na}}, \bibinfo {author} {\bibfnamefont
  {A.}~\bibnamefont {Clairon}}, \bibinfo {author} {\bibfnamefont
  {M.}~\bibnamefont {Tobar}}, \bibinfo {author} {\bibfnamefont
  {S.}~\bibnamefont {Bize}}, \bibinfo {author} {\bibfnamefont {Y.}~\bibnamefont
  {Le~Coq}}, \ and\ \bibinfo {author} {\bibfnamefont {G.}~\bibnamefont
  {Santarelli}},\ }\href@noop {} {\bibfield  {journal} {\bibinfo  {journal}
  {Appl. Phys. Lett.}\ }\textbf {\bibinfo {volume} {94}},\ \bibinfo {pages}
  {141105} (\bibinfo {year} {2009}{\natexlab{b}})}\BibitemShut {NoStop}%
\bibitem [{\citenamefont {Zhang}\ \emph {et~al.}(2012)\citenamefont {Zhang},
  \citenamefont {Li}, \citenamefont {Lours}, \citenamefont {Seidelin},
  \citenamefont {Santarelli},\ and\ \citenamefont {Le~Coq}}]{ZhangAPB2012}%
  \BibitemOpen
  \bibfield  {author} {\bibinfo {author} {\bibfnamefont {W.}~\bibnamefont
  {Zhang}}, \bibinfo {author} {\bibfnamefont {T.}~\bibnamefont {Li}}, \bibinfo
  {author} {\bibfnamefont {M.}~\bibnamefont {Lours}}, \bibinfo {author}
  {\bibfnamefont {S.}~\bibnamefont {Seidelin}}, \bibinfo {author}
  {\bibfnamefont {G.}~\bibnamefont {Santarelli}}, \ and\ \bibinfo {author}
  {\bibfnamefont {Y.}~\bibnamefont {Le~Coq}},\ }\href@noop {} {\bibfield
  {journal} {\bibinfo  {journal} {Appl. Phys. B Lasers and Optics}\ }\textbf
  {\bibinfo {volume} {106}},\ \bibinfo {pages} {301} (\bibinfo {year}
  {2012})}\BibitemShut {NoStop}%
\bibitem [{\citenamefont {Haboucha}\ \emph {et~al.}(2011)\citenamefont
  {Haboucha}, \citenamefont {Zhang}, \citenamefont {Li}, \citenamefont {Lours},
  \citenamefont {Luiten}, \citenamefont {Le~Coq},\ and\ \citenamefont
  {Santarelli}}]{HabouchaOL2011}%
  \BibitemOpen
  \bibfield  {author} {\bibinfo {author} {\bibfnamefont {A.}~\bibnamefont
  {Haboucha}}, \bibinfo {author} {\bibfnamefont {W.}~\bibnamefont {Zhang}},
  \bibinfo {author} {\bibfnamefont {T.}~\bibnamefont {Li}}, \bibinfo {author}
  {\bibfnamefont {M.}~\bibnamefont {Lours}}, \bibinfo {author} {\bibfnamefont
  {A.}~\bibnamefont {Luiten}}, \bibinfo {author} {\bibfnamefont
  {Y.}~\bibnamefont {Le~Coq}}, \ and\ \bibinfo {author} {\bibfnamefont
  {G.}~\bibnamefont {Santarelli}},\ }\href@noop {} {\bibfield  {journal}
  {\bibinfo  {journal} {Opt. Lett.}\ }\textbf {\bibinfo {volume} {36}},\
  \bibinfo {pages} {3654} (\bibinfo {year} {2011})}\BibitemShut {NoStop}%
\bibitem [{\citenamefont {Zhang}\ \emph {et~al.}(2014)\citenamefont {Zhang},
  \citenamefont {Seidelin}, \citenamefont {Joshi}, \citenamefont {Datta},
  \citenamefont {Santarelli},\ and\ \citenamefont {Le~Coq}}]{ZhangOL2014}%
  \BibitemOpen
  \bibfield  {author} {\bibinfo {author} {\bibfnamefont {W.}~\bibnamefont
  {Zhang}}, \bibinfo {author} {\bibfnamefont {S.}~\bibnamefont {Seidelin}},
  \bibinfo {author} {\bibfnamefont {A.}~\bibnamefont {Joshi}}, \bibinfo
  {author} {\bibfnamefont {S.}~\bibnamefont {Datta}}, \bibinfo {author}
  {\bibfnamefont {G.}~\bibnamefont {Santarelli}}, \ and\ \bibinfo {author}
  {\bibfnamefont {Y.}~\bibnamefont {Le~Coq}},\ }\href@noop {} {\bibfield
  {journal} {\bibinfo  {journal} {Opt. Lett.}\ }\textbf {\bibinfo {volume}
  {39}},\ \bibinfo {pages} {1204} (\bibinfo {year} {2014})}\BibitemShut
  {NoStop}%
\bibitem [{\citenamefont {Nicolodi}\ \emph {et~al.}(2014)\citenamefont
  {Nicolodi}, \citenamefont {Argence}, \citenamefont {Zhang}, \citenamefont
  {Le~Targat}, \citenamefont {Santarelli},\ and\ \citenamefont
  {Le~Coq}}]{NicolodiNP2014}%
  \BibitemOpen
  \bibfield  {author} {\bibinfo {author} {\bibfnamefont {D.}~\bibnamefont
  {Nicolodi}}, \bibinfo {author} {\bibfnamefont {B.}~\bibnamefont {Argence}},
  \bibinfo {author} {\bibfnamefont {W.}~\bibnamefont {Zhang}}, \bibinfo
  {author} {\bibfnamefont {R.}~\bibnamefont {Le~Targat}}, \bibinfo {author}
  {\bibfnamefont {G.}~\bibnamefont {Santarelli}}, \ and\ \bibinfo {author}
  {\bibfnamefont {Y.}~\bibnamefont {Le~Coq}},\ }\href@noop {} {\bibfield
  {journal} {\bibinfo  {journal} {Nature Photonics}\ }\textbf {\bibinfo
  {volume} {8}},\ \bibinfo {pages} {219} (\bibinfo {year} {2014})}\BibitemShut
  {NoStop}%
\bibitem [{\citenamefont {Gu\'ena}\ \emph {et~al.}(2012)\citenamefont
  {Gu\'ena}, \citenamefont {Abgrall}, \citenamefont {Rovera}, \citenamefont
  {Rosenbusch}, \citenamefont {Tobar}, \citenamefont {Laurent}, \citenamefont
  {Clairon},\ and\ \citenamefont {Bize}}]{guena2012b}%
  \BibitemOpen
  \bibfield  {author} {\bibinfo {author} {\bibfnamefont {J.}~\bibnamefont
  {Gu\'ena}}, \bibinfo {author} {\bibfnamefont {M.}~\bibnamefont {Abgrall}},
  \bibinfo {author} {\bibfnamefont {D.}~\bibnamefont {Rovera}}, \bibinfo
  {author} {\bibfnamefont {P.}~\bibnamefont {Rosenbusch}}, \bibinfo {author}
  {\bibfnamefont {M.~E.}\ \bibnamefont {Tobar}}, \bibinfo {author}
  {\bibfnamefont {P.}~\bibnamefont {Laurent}}, \bibinfo {author} {\bibfnamefont
  {A.}~\bibnamefont {Clairon}}, \ and\ \bibinfo {author} {\bibfnamefont
  {S.}~\bibnamefont {Bize}},\ }\href@noop {} {\bibfield  {journal} {\bibinfo
  {journal} {Phys. Rev. Lett.}\ }\textbf {\bibinfo {volume} {109}},\ \bibinfo
  {pages} {080801} (\bibinfo {year} {2012})}\BibitemShut {NoStop}%
\bibitem [{\citenamefont {Peil}\ \emph {et~al.}(2013)\citenamefont {Peil},
  \citenamefont {Crane}, \citenamefont {Hanssen}, \citenamefont {Swanson},\
  and\ \citenamefont {Ekstrom}}]{peil2013}%
  \BibitemOpen
  \bibfield  {author} {\bibinfo {author} {\bibfnamefont {S.}~\bibnamefont
  {Peil}}, \bibinfo {author} {\bibfnamefont {S.}~\bibnamefont {Crane}},
  \bibinfo {author} {\bibfnamefont {J.~L.}\ \bibnamefont {Hanssen}}, \bibinfo
  {author} {\bibfnamefont {T.~B.}\ \bibnamefont {Swanson}}, \ and\ \bibinfo
  {author} {\bibfnamefont {C.~R.}\ \bibnamefont {Ekstrom}},\ }\href@noop {}
  {\bibfield  {journal} {\bibinfo  {journal} {Phys. Rev. A}\ }\textbf {\bibinfo
  {volume} {87}},\ \bibinfo {pages} {010102} (\bibinfo {year}
  {2013})}\BibitemShut {NoStop}%
\bibitem [{CCT(2012)}]{CCTF2012}%
  \BibitemOpen
  \href@noop {} {\bibfield  {journal} {\bibinfo  {journal} {Report of the
  19$^{\mathrm{th}}$ meeting (13-14 September 2012) to the International
  Committee for Weights and Measures}\ ,\ \bibinfo {pages} {p.59}} (\bibinfo
  {year} {2012})}\BibitemShut {NoStop}%
\bibitem [{\citenamefont {Fischer}\ \emph {et~al.}(2004)\citenamefont
  {Fischer}, \citenamefont {Kolachevsky}, \citenamefont {Zimmermann},
  \citenamefont {Holzwarth}, \citenamefont {Udem}, \citenamefont {H{\"a}nsch},
  \citenamefont {Abgrall}, \citenamefont {Gr{\"u}nert}, \citenamefont
  {Maksimovic}, \citenamefont {Bize} \emph {et~al.}}]{fischer2004}%
  \BibitemOpen
  \bibfield  {author} {\bibinfo {author} {\bibfnamefont {M.}~\bibnamefont
  {Fischer}}, \bibinfo {author} {\bibfnamefont {N.}~\bibnamefont
  {Kolachevsky}}, \bibinfo {author} {\bibfnamefont {M.}~\bibnamefont
  {Zimmermann}}, \bibinfo {author} {\bibfnamefont {R.}~\bibnamefont
  {Holzwarth}}, \bibinfo {author} {\bibfnamefont {T.}~\bibnamefont {Udem}},
  \bibinfo {author} {\bibfnamefont {T.}~\bibnamefont {H{\"a}nsch}}, \bibinfo
  {author} {\bibfnamefont {M.}~\bibnamefont {Abgrall}}, \bibinfo {author}
  {\bibfnamefont {J.}~\bibnamefont {Gr{\"u}nert}}, \bibinfo {author}
  {\bibfnamefont {I.}~\bibnamefont {Maksimovic}}, \bibinfo {author}
  {\bibfnamefont {S.}~\bibnamefont {Bize}},  \emph {et~al.},\ }\href@noop {}
  {\bibfield  {journal} {\bibinfo  {journal} {Phys. Rev. Lett.}\ }\textbf
  {\bibinfo {volume} {92}},\ \bibinfo {pages} {230802} (\bibinfo {year}
  {2004})}\BibitemShut {NoStop}%
\bibitem [{\citenamefont {Tamm}\ \emph {et~al.}(2014)\citenamefont {Tamm},
  \citenamefont {Huntemann}, \citenamefont {Lipphardt}, \citenamefont
  {Gerginov}, \citenamefont {Nemitz}, \citenamefont {Kazda}, \citenamefont
  {Weyers},\ and\ \citenamefont {Peik}}]{tamm2014}%
  \BibitemOpen
  \bibfield  {author} {\bibinfo {author} {\bibfnamefont {C.}~\bibnamefont
  {Tamm}}, \bibinfo {author} {\bibfnamefont {N.}~\bibnamefont {Huntemann}},
  \bibinfo {author} {\bibfnamefont {B.}~\bibnamefont {Lipphardt}}, \bibinfo
  {author} {\bibfnamefont {V.}~\bibnamefont {Gerginov}}, \bibinfo {author}
  {\bibfnamefont {N.}~\bibnamefont {Nemitz}}, \bibinfo {author} {\bibfnamefont
  {M.}~\bibnamefont {Kazda}}, \bibinfo {author} {\bibfnamefont
  {S.}~\bibnamefont {Weyers}}, \ and\ \bibinfo {author} {\bibfnamefont
  {E.}~\bibnamefont {Peik}},\ }\href@noop {} {\bibfield  {journal} {\bibinfo
  {journal} {Phys. Rev. A}\ }\textbf {\bibinfo {volume} {89}},\ \bibinfo
  {pages} {023820} (\bibinfo {year} {2014})}\BibitemShut {NoStop}%
\bibitem [{\citenamefont {Fortier}\ \emph {et~al.}(2007)\citenamefont
  {Fortier}, \citenamefont {Ashby}, \citenamefont {Bergquist}, \citenamefont
  {Delaney}, \citenamefont {Diddams}, \citenamefont {Heavner}, \citenamefont
  {Hollberg}, \citenamefont {Itano}, \citenamefont {Jefferts}, \citenamefont
  {Kim} \emph {et~al.}}]{fortier2007}%
  \BibitemOpen
  \bibfield  {author} {\bibinfo {author} {\bibfnamefont {T.}~\bibnamefont
  {Fortier}}, \bibinfo {author} {\bibfnamefont {N.}~\bibnamefont {Ashby}},
  \bibinfo {author} {\bibfnamefont {J.}~\bibnamefont {Bergquist}}, \bibinfo
  {author} {\bibfnamefont {M.}~\bibnamefont {Delaney}}, \bibinfo {author}
  {\bibfnamefont {S.}~\bibnamefont {Diddams}}, \bibinfo {author} {\bibfnamefont
  {T.}~\bibnamefont {Heavner}}, \bibinfo {author} {\bibfnamefont
  {L.}~\bibnamefont {Hollberg}}, \bibinfo {author} {\bibfnamefont
  {W.}~\bibnamefont {Itano}}, \bibinfo {author} {\bibfnamefont
  {S.}~\bibnamefont {Jefferts}}, \bibinfo {author} {\bibfnamefont
  {K.}~\bibnamefont {Kim}},  \emph {et~al.},\ }\href@noop {} {\bibfield
  {journal} {\bibinfo  {journal} {Phys. Rev. Lett.}\ }\textbf {\bibinfo
  {volume} {98}},\ \bibinfo {pages} {070801} (\bibinfo {year}
  {2007})}\BibitemShut {NoStop}%
\bibitem [{\citenamefont {Leefer}\ \emph {et~al.}(2013)\citenamefont {Leefer},
  \citenamefont {Weber}, \citenamefont {Cing{\"o}z}, \citenamefont
  {Torgerson},\ and\ \citenamefont {Budker}}]{leefer2013}%
  \BibitemOpen
  \bibfield  {author} {\bibinfo {author} {\bibfnamefont {N.}~\bibnamefont
  {Leefer}}, \bibinfo {author} {\bibfnamefont {C.}~\bibnamefont {Weber}},
  \bibinfo {author} {\bibfnamefont {A.}~\bibnamefont {Cing{\"o}z}}, \bibinfo
  {author} {\bibfnamefont {J.}~\bibnamefont {Torgerson}}, \ and\ \bibinfo
  {author} {\bibfnamefont {D.}~\bibnamefont {Budker}},\ }\href@noop {}
  {\bibfield  {journal} {\bibinfo  {journal} {Phys. Rev. Lett.}\ }\textbf
  {\bibinfo {volume} {111}},\ \bibinfo {pages} {060801} (\bibinfo {year}
  {2013})}\BibitemShut {NoStop}%
\bibitem [{\citenamefont {Rosenband}\ \emph {et~al.}(2008)\citenamefont
  {Rosenband}, \citenamefont {Hume}, \citenamefont {Schmidt}, \citenamefont
  {Chou}, \citenamefont {Brusch}, \citenamefont {Lorini}, \citenamefont
  {Oskay}, \citenamefont {Drullinger}, \citenamefont {Fortier}, \citenamefont
  {Stalnaker} \emph {et~al.}}]{rosenband2008}%
  \BibitemOpen
  \bibfield  {author} {\bibinfo {author} {\bibfnamefont {T.}~\bibnamefont
  {Rosenband}}, \bibinfo {author} {\bibfnamefont {D.}~\bibnamefont {Hume}},
  \bibinfo {author} {\bibfnamefont {P.}~\bibnamefont {Schmidt}}, \bibinfo
  {author} {\bibfnamefont {C.}~\bibnamefont {Chou}}, \bibinfo {author}
  {\bibfnamefont {A.}~\bibnamefont {Brusch}}, \bibinfo {author} {\bibfnamefont
  {L.}~\bibnamefont {Lorini}}, \bibinfo {author} {\bibfnamefont
  {W.}~\bibnamefont {Oskay}}, \bibinfo {author} {\bibfnamefont
  {R.}~\bibnamefont {Drullinger}}, \bibinfo {author} {\bibfnamefont
  {T.}~\bibnamefont {Fortier}}, \bibinfo {author} {\bibfnamefont
  {J.}~\bibnamefont {Stalnaker}},  \emph {et~al.},\ }\href@noop {} {\bibfield
  {journal} {\bibinfo  {journal} {Science}\ }\textbf {\bibinfo {volume}
  {319}},\ \bibinfo {pages} {1808} (\bibinfo {year} {2008})}\BibitemShut
  {NoStop}%
\bibitem [{\citenamefont {Madej}\ \emph {et~al.}(2012)\citenamefont {Madej},
  \citenamefont {Dub\'e}, \citenamefont {Zhou}, \citenamefont {Bernard},\ and\
  \citenamefont {Gertsvolf}}]{Madej2012}%
  \BibitemOpen
  \bibfield  {author} {\bibinfo {author} {\bibfnamefont {A.~A.}\ \bibnamefont
  {Madej}}, \bibinfo {author} {\bibfnamefont {P.}~\bibnamefont {Dub\'e}},
  \bibinfo {author} {\bibfnamefont {Z.}~\bibnamefont {Zhou}}, \bibinfo {author}
  {\bibfnamefont {J.~E.}\ \bibnamefont {Bernard}}, \ and\ \bibinfo {author}
  {\bibfnamefont {M.}~\bibnamefont {Gertsvolf}},\ }\href@noop {} {\bibfield
  {journal} {\bibinfo  {journal} {Phys. Rev. Lett.}\ }\textbf {\bibinfo
  {volume} {109}},\ \bibinfo {pages} {203002} (\bibinfo {year}
  {2012})}\BibitemShut {NoStop}%
\bibitem [{\citenamefont {Barwood}\ \emph {et~al.}(2014)\citenamefont
  {Barwood}, \citenamefont {Huang}, \citenamefont {Klein}, \citenamefont
  {Johnson}, \citenamefont {King}, \citenamefont {Margolis}, \citenamefont
  {Szymaniec},\ and\ \citenamefont {Gill}}]{Barwood2014}%
  \BibitemOpen
  \bibfield  {author} {\bibinfo {author} {\bibfnamefont {G.~P.}\ \bibnamefont
  {Barwood}}, \bibinfo {author} {\bibfnamefont {G.}~\bibnamefont {Huang}},
  \bibinfo {author} {\bibfnamefont {H.~A.}\ \bibnamefont {Klein}}, \bibinfo
  {author} {\bibfnamefont {L.~A.~M.}\ \bibnamefont {Johnson}}, \bibinfo
  {author} {\bibfnamefont {S.~A.}\ \bibnamefont {King}}, \bibinfo {author}
  {\bibfnamefont {H.~S.}\ \bibnamefont {Margolis}}, \bibinfo {author}
  {\bibfnamefont {K.}~\bibnamefont {Szymaniec}}, \ and\ \bibinfo {author}
  {\bibfnamefont {P.}~\bibnamefont {Gill}},\ }\href@noop {} {\bibfield
  {journal} {\bibinfo  {journal} {Phys. Rev. A}\ }\textbf {\bibinfo {volume}
  {89}},\ \bibinfo {pages} {050501} (\bibinfo {year} {2014})}\BibitemShut
  {NoStop}%
\bibitem [{\citenamefont {Lemke}\ \emph {et~al.}(2009)\citenamefont {Lemke},
  \citenamefont {Ludlow}, \citenamefont {Barber}, \citenamefont {Fortier},
  \citenamefont {Diddams}, \citenamefont {Jiang}, \citenamefont {Jefferts},
  \citenamefont {Heavner}, \citenamefont {Parker},\ and\ \citenamefont
  {Oates}}]{Lemke2009a}%
  \BibitemOpen
  \bibfield  {author} {\bibinfo {author} {\bibfnamefont {N.~D.}\ \bibnamefont
  {Lemke}}, \bibinfo {author} {\bibfnamefont {A.~D.}\ \bibnamefont {Ludlow}},
  \bibinfo {author} {\bibfnamefont {Z.~W.}\ \bibnamefont {Barber}}, \bibinfo
  {author} {\bibfnamefont {T.~M.}\ \bibnamefont {Fortier}}, \bibinfo {author}
  {\bibfnamefont {S.~A.}\ \bibnamefont {Diddams}}, \bibinfo {author}
  {\bibfnamefont {Y.}~\bibnamefont {Jiang}}, \bibinfo {author} {\bibfnamefont
  {S.~R.}\ \bibnamefont {Jefferts}}, \bibinfo {author} {\bibfnamefont {T.~P.}\
  \bibnamefont {Heavner}}, \bibinfo {author} {\bibfnamefont {T.~E.}\
  \bibnamefont {Parker}}, \ and\ \bibinfo {author} {\bibfnamefont {C.~W.}\
  \bibnamefont {Oates}},\ }\href@noop {} {\bibfield  {journal} {\bibinfo
  {journal} {Phys. Rev. Lett.}\ }\textbf {\bibinfo {volume} {103}},\ \bibinfo
  {pages} {063001} (\bibinfo {year} {2009})}\BibitemShut {NoStop}%
\bibitem [{\citenamefont {Petit}\ \emph {et~al.}(2015)\citenamefont {Petit},
  \citenamefont {Arias},\ and\ \citenamefont {Panfilo}}]{cras2014Petit}%
  \BibitemOpen
  \bibfield  {author} {\bibinfo {author} {\bibfnamefont {G.}~\bibnamefont
  {Petit}}, \bibinfo {author} {\bibfnamefont {F.}~\bibnamefont {Arias}}, \ and\
  \bibinfo {author} {\bibfnamefont {G.}~\bibnamefont {Panfilo}},\ }\href
  {\doibase http://dx.doi.org/10.1016/j.crhy.2015.03.002} {\bibfield  {journal}
  {\bibinfo  {journal} {Comptes Rendus Physique}\ }\textbf {\bibinfo {volume}
  {16}},\ \bibinfo {pages} {480 } (\bibinfo {year} {2015})},\ \bibinfo {note}
  {the measurement of time / La mesure du temps}\BibitemShut {NoStop}%
\bibitem [{\citenamefont {Laurent}\ \emph {et~al.}(2006)\citenamefont
  {Laurent}, \citenamefont {Abgrall}, \citenamefont {Jentsch}, \citenamefont
  {Lemonde}, \citenamefont {Santarelli}, \citenamefont {Clairon}, \citenamefont
  {Maksimovic}, \citenamefont {Bize}, \citenamefont {Salomon}, \citenamefont
  {Blonde} \emph {et~al.}}]{laurent2006}%
  \BibitemOpen
  \bibfield  {author} {\bibinfo {author} {\bibfnamefont {P.}~\bibnamefont
  {Laurent}}, \bibinfo {author} {\bibfnamefont {M.}~\bibnamefont {Abgrall}},
  \bibinfo {author} {\bibfnamefont {C.}~\bibnamefont {Jentsch}}, \bibinfo
  {author} {\bibfnamefont {P.}~\bibnamefont {Lemonde}}, \bibinfo {author}
  {\bibfnamefont {G.}~\bibnamefont {Santarelli}}, \bibinfo {author}
  {\bibfnamefont {A.}~\bibnamefont {Clairon}}, \bibinfo {author} {\bibfnamefont
  {I.}~\bibnamefont {Maksimovic}}, \bibinfo {author} {\bibfnamefont
  {S.}~\bibnamefont {Bize}}, \bibinfo {author} {\bibfnamefont {C.}~\bibnamefont
  {Salomon}}, \bibinfo {author} {\bibfnamefont {D.}~\bibnamefont {Blonde}},
  \emph {et~al.},\ }\href@noop {} {\bibfield  {journal} {\bibinfo  {journal}
  {Appl. Phys. B}\ }\textbf {\bibinfo {volume} {84}},\ \bibinfo {pages} {683}
  (\bibinfo {year} {2006})}\BibitemShut {NoStop}%
\bibitem [{\citenamefont {Cacciapuoti}\ \emph {et~al.}(2007)\citenamefont
  {Cacciapuoti}, \citenamefont {Dimarcq}, \citenamefont {Santarelli},
  \citenamefont {Laurent}, \citenamefont {Lemonde}, \citenamefont {Clairon},
  \citenamefont {Berthoud}, \citenamefont {Jornod}, \citenamefont {Reina},
  \citenamefont {Feltham} \emph {et~al.}}]{cacciapuoti2007}%
  \BibitemOpen
  \bibfield  {author} {\bibinfo {author} {\bibfnamefont {L.}~\bibnamefont
  {Cacciapuoti}}, \bibinfo {author} {\bibfnamefont {N.}~\bibnamefont
  {Dimarcq}}, \bibinfo {author} {\bibfnamefont {G.}~\bibnamefont {Santarelli}},
  \bibinfo {author} {\bibfnamefont {P.}~\bibnamefont {Laurent}}, \bibinfo
  {author} {\bibfnamefont {P.}~\bibnamefont {Lemonde}}, \bibinfo {author}
  {\bibfnamefont {A.}~\bibnamefont {Clairon}}, \bibinfo {author} {\bibfnamefont
  {P.}~\bibnamefont {Berthoud}}, \bibinfo {author} {\bibfnamefont
  {A.}~\bibnamefont {Jornod}}, \bibinfo {author} {\bibfnamefont
  {F.}~\bibnamefont {Reina}}, \bibinfo {author} {\bibfnamefont
  {S.}~\bibnamefont {Feltham}},  \emph {et~al.},\ }\href@noop {} {\bibfield
  {journal} {\bibinfo  {journal} {Nuclear Physics B-Proceedings Supplements}\
  }\textbf {\bibinfo {volume} {166}},\ \bibinfo {pages} {303} (\bibinfo {year}
  {2007})}\BibitemShut {NoStop}%
\bibitem [{\citenamefont {Cacciapuoti}\ and\ \citenamefont
  {Salomon}(2009)}]{cacciapuoti2009}%
  \BibitemOpen
  \bibfield  {author} {\bibinfo {author} {\bibfnamefont {L.}~\bibnamefont
  {Cacciapuoti}}\ and\ \bibinfo {author} {\bibfnamefont {C.}~\bibnamefont
  {Salomon}},\ }\href@noop {} {\bibfield  {journal} {\bibinfo  {journal} {The
  European Physical Journal Special Topics}\ }\textbf {\bibinfo {volume}
  {172}},\ \bibinfo {pages} {57} (\bibinfo {year} {2009})}\BibitemShut
  {NoStop}%
\bibitem [{\citenamefont {Gu{\'e}na}\ \emph {et~al.}(2014)\citenamefont
  {Gu{\'e}na}, \citenamefont {Abgrall}, \citenamefont {Clairon},\ and\
  \citenamefont {Bize}}]{guena2014}%
  \BibitemOpen
  \bibfield  {author} {\bibinfo {author} {\bibfnamefont {J.}~\bibnamefont
  {Gu{\'e}na}}, \bibinfo {author} {\bibfnamefont {M.}~\bibnamefont {Abgrall}},
  \bibinfo {author} {\bibfnamefont {A.}~\bibnamefont {Clairon}}, \ and\
  \bibinfo {author} {\bibfnamefont {S.}~\bibnamefont {Bize}},\ }\href@noop {}
  {\bibfield  {journal} {\bibinfo  {journal} {Metrologia}\ }\textbf {\bibinfo
  {volume} {51}},\ \bibinfo {pages} {108} (\bibinfo {year} {2014})}\BibitemShut
  {NoStop}%
\bibitem [{\citenamefont {Rovera}\ \emph {et~al.}(2013)\citenamefont {Rovera},
  \citenamefont {Abgrall}, \citenamefont {Bize}, \citenamefont {Chupin},
  \citenamefont {Gu\'ena}, \citenamefont {Laurent}, \citenamefont
  {Rosenbusch},\ and\ \citenamefont {Uhrich}}]{rovera2013}%
  \BibitemOpen
  \bibfield  {author} {\bibinfo {author} {\bibfnamefont {G.}~\bibnamefont
  {Rovera}}, \bibinfo {author} {\bibfnamefont {M.}~\bibnamefont {Abgrall}},
  \bibinfo {author} {\bibfnamefont {S.}~\bibnamefont {Bize}}, \bibinfo {author}
  {\bibfnamefont {B.}~\bibnamefont {Chupin}}, \bibinfo {author} {\bibfnamefont
  {J.}~\bibnamefont {Gu\'ena}}, \bibinfo {author} {\bibfnamefont
  {P.}~\bibnamefont {Laurent}}, \bibinfo {author} {\bibfnamefont
  {P.}~\bibnamefont {Rosenbusch}}, \ and\ \bibinfo {author} {\bibfnamefont
  {P.}~\bibnamefont {Uhrich}},\ }in\ \href@noop {} {\emph {\bibinfo {booktitle}
  {proceedings of the 27$^{\mathrm{th}}$ {European Frequency and Time Forum
  joint with the 2013 IEEE International Frequency Control Symposium}}}}\
  (\bibinfo {year} {2013})\ pp.\ \bibinfo {pages} {649--651}\BibitemShut
  {NoStop}%
\bibitem [{\citenamefont {Abgrall}\ \emph {et~al.}(2014)\citenamefont
  {Abgrall}, \citenamefont {Bize}, \citenamefont {Chupin}, \citenamefont
  {Gu{\'e}na}, \citenamefont {Laurent}, \citenamefont {Rosenbusch},
  \citenamefont {Uhrich},\ and\ \citenamefont {Rovera}}]{abgrall2014}%
  \BibitemOpen
  \bibfield  {author} {\bibinfo {author} {\bibfnamefont {M.}~\bibnamefont
  {Abgrall}}, \bibinfo {author} {\bibfnamefont {S.}~\bibnamefont {Bize}},
  \bibinfo {author} {\bibfnamefont {B.}~\bibnamefont {Chupin}}, \bibinfo
  {author} {\bibfnamefont {J.}~\bibnamefont {Gu{\'e}na}}, \bibinfo {author}
  {\bibfnamefont {P.}~\bibnamefont {Laurent}}, \bibinfo {author} {\bibfnamefont
  {P.}~\bibnamefont {Rosenbusch}}, \bibinfo {author} {\bibfnamefont
  {P.}~\bibnamefont {Uhrich}}, \ and\ \bibinfo {author} {\bibfnamefont {G.~D.}\
  \bibnamefont {Rovera}},\ }in\ \href@noop {} {\emph {\bibinfo {booktitle}
  {proceedings. of the 28$^{\mathrm{th}}$ {European Frequency and Time
  Forum}}}}\ (\bibinfo {year} {2014})\ p.\ \bibinfo {pages} {564}\BibitemShut
  {NoStop}%
\bibitem [{\citenamefont {Baillard}\ \emph {et~al.}(2008)\citenamefont
  {Baillard}, \citenamefont {Fouch\'{e}}, \citenamefont {Le~Targat},
  \citenamefont {Westergaard}, \citenamefont {Lecallier}, \citenamefont
  {Chapelet}, \citenamefont {Abgrall}, \citenamefont {Rovera}, \citenamefont
  {Laurent}, \citenamefont {Rosenbusch}, \citenamefont {Bize}, \citenamefont
  {Santarelli}, \citenamefont {Clairon}, \citenamefont {Lemonde}, \citenamefont
  {Grosche}, \citenamefont {Lipphardt},\ and\ \citenamefont
  {Schnatz}}]{Baillard2007b}%
  \BibitemOpen
  \bibfield  {author} {\bibinfo {author} {\bibfnamefont {X.}~\bibnamefont
  {Baillard}}, \bibinfo {author} {\bibfnamefont {M.}~\bibnamefont
  {Fouch\'{e}}}, \bibinfo {author} {\bibfnamefont {R.}~\bibnamefont
  {Le~Targat}}, \bibinfo {author} {\bibfnamefont {P.~G.}\ \bibnamefont
  {Westergaard}}, \bibinfo {author} {\bibfnamefont {A.}~\bibnamefont
  {Lecallier}}, \bibinfo {author} {\bibfnamefont {F.}~\bibnamefont {Chapelet}},
  \bibinfo {author} {\bibfnamefont {M.}~\bibnamefont {Abgrall}}, \bibinfo
  {author} {\bibfnamefont {G.~D.}\ \bibnamefont {Rovera}}, \bibinfo {author}
  {\bibfnamefont {P.}~\bibnamefont {Laurent}}, \bibinfo {author} {\bibfnamefont
  {P.}~\bibnamefont {Rosenbusch}}, \bibinfo {author} {\bibfnamefont
  {S.}~\bibnamefont {Bize}}, \bibinfo {author} {\bibfnamefont {G.}~\bibnamefont
  {Santarelli}}, \bibinfo {author} {\bibfnamefont {A.}~\bibnamefont {Clairon}},
  \bibinfo {author} {\bibfnamefont {P.}~\bibnamefont {Lemonde}}, \bibinfo
  {author} {\bibfnamefont {G.}~\bibnamefont {Grosche}}, \bibinfo {author}
  {\bibfnamefont {B.}~\bibnamefont {Lipphardt}}, \ and\ \bibinfo {author}
  {\bibfnamefont {H.}~\bibnamefont {Schnatz}},\ }\href@noop {} {\bibfield
  {journal} {\bibinfo  {journal} {Eur. Phys. J. D}\ }\textbf {\bibinfo {volume}
  {48}},\ \bibinfo {pages} {11} (\bibinfo {year} {2008})}\BibitemShut {NoStop}%
\bibitem [{\citenamefont {Baillard}\ \emph {et~al.}(2007)\citenamefont
  {Baillard}, \citenamefont {Fouch\'{e}}, \citenamefont {Le~Targat},
  \citenamefont {Westergaard}, \citenamefont {Lecallier}, \citenamefont
  {Le~Coq}, \citenamefont {Rovera}, \citenamefont {Bize},\ and\ \citenamefont
  {Lemonde}}]{Baillard2007}%
  \BibitemOpen
  \bibfield  {author} {\bibinfo {author} {\bibfnamefont {X.}~\bibnamefont
  {Baillard}}, \bibinfo {author} {\bibfnamefont {M.}~\bibnamefont
  {Fouch\'{e}}}, \bibinfo {author} {\bibfnamefont {R.}~\bibnamefont
  {Le~Targat}}, \bibinfo {author} {\bibfnamefont {P.~G.}\ \bibnamefont
  {Westergaard}}, \bibinfo {author} {\bibfnamefont {A.}~\bibnamefont
  {Lecallier}}, \bibinfo {author} {\bibfnamefont {Y.}~\bibnamefont {Le~Coq}},
  \bibinfo {author} {\bibfnamefont {G.~D.}\ \bibnamefont {Rovera}}, \bibinfo
  {author} {\bibfnamefont {S.}~\bibnamefont {Bize}}, \ and\ \bibinfo {author}
  {\bibfnamefont {P.}~\bibnamefont {Lemonde}},\ }\href@noop {} {\bibfield
  {journal} {\bibinfo  {journal} {Opt. Lett.}\ }\textbf {\bibinfo {volume}
  {32}},\ \bibinfo {pages} {1812} (\bibinfo {year} {2007})}\BibitemShut
  {NoStop}%
\bibitem [{\citenamefont {Parthey}\ \emph {et~al.}(2011)\citenamefont
  {Parthey}, \citenamefont {Matveev}, \citenamefont {Alnis}, \citenamefont
  {Bernhardt}, \citenamefont {Beyer}, \citenamefont {Holzwarth}, \citenamefont
  {Maistrou}, \citenamefont {Pohl}, \citenamefont {Predehl}, \citenamefont
  {Udem} \emph {et~al.}}]{parthey2011}%
  \BibitemOpen
  \bibfield  {author} {\bibinfo {author} {\bibfnamefont {C.~G.}\ \bibnamefont
  {Parthey}}, \bibinfo {author} {\bibfnamefont {A.}~\bibnamefont {Matveev}},
  \bibinfo {author} {\bibfnamefont {J.}~\bibnamefont {Alnis}}, \bibinfo
  {author} {\bibfnamefont {B.}~\bibnamefont {Bernhardt}}, \bibinfo {author}
  {\bibfnamefont {A.}~\bibnamefont {Beyer}}, \bibinfo {author} {\bibfnamefont
  {R.}~\bibnamefont {Holzwarth}}, \bibinfo {author} {\bibfnamefont
  {A.}~\bibnamefont {Maistrou}}, \bibinfo {author} {\bibfnamefont
  {R.}~\bibnamefont {Pohl}}, \bibinfo {author} {\bibfnamefont {K.}~\bibnamefont
  {Predehl}}, \bibinfo {author} {\bibfnamefont {T.}~\bibnamefont {Udem}},
  \emph {et~al.},\ }\href@noop {} {\bibfield  {journal} {\bibinfo  {journal}
  {Phys. Rev. Lett.}\ }\textbf {\bibinfo {volume} {107}},\ \bibinfo {pages}
  {203001} (\bibinfo {year} {2011})}\BibitemShut {NoStop}%
\bibitem [{\citenamefont {Chwalla}\ \emph {et~al.}(2009)\citenamefont
  {Chwalla}, \citenamefont {Benhelm}, \citenamefont {Kim}, \citenamefont
  {Kirchmair}, \citenamefont {Monz}, \citenamefont {Riebe}, \citenamefont
  {Schindler}, \citenamefont {Villar}, \citenamefont {H{\"a}nsel},
  \citenamefont {Roos} \emph {et~al.}}]{chwalla2009}%
  \BibitemOpen
  \bibfield  {author} {\bibinfo {author} {\bibfnamefont {M.}~\bibnamefont
  {Chwalla}}, \bibinfo {author} {\bibfnamefont {J.}~\bibnamefont {Benhelm}},
  \bibinfo {author} {\bibfnamefont {K.}~\bibnamefont {Kim}}, \bibinfo {author}
  {\bibfnamefont {G.}~\bibnamefont {Kirchmair}}, \bibinfo {author}
  {\bibfnamefont {T.}~\bibnamefont {Monz}}, \bibinfo {author} {\bibfnamefont
  {M.}~\bibnamefont {Riebe}}, \bibinfo {author} {\bibfnamefont
  {P.}~\bibnamefont {Schindler}}, \bibinfo {author} {\bibfnamefont
  {A.}~\bibnamefont {Villar}}, \bibinfo {author} {\bibfnamefont
  {W.}~\bibnamefont {H{\"a}nsel}}, \bibinfo {author} {\bibfnamefont
  {C.}~\bibnamefont {Roos}},  \emph {et~al.},\ }\href@noop {} {\bibfield
  {journal} {\bibinfo  {journal} {Phys. Rev. Lett.}\ }\textbf {\bibinfo
  {volume} {102}},\ \bibinfo {pages} {023002} (\bibinfo {year}
  {2009})}\BibitemShut {NoStop}%
\end{thebibliography}%

\end{document}